\documentclass[12pt,preprint,eqsecnum,aps]{revtex4}

\usepackage{amssymb,amsmath,mathrsfs,epsfig}

\newcommand{\rf}[1]{(\,\ref{#1}\,)}
\newcommand{\txt}[1]{\textstyle{#1}}
\newcommand{\bs}{\,-\,}
\newcommand{\kl}{(\,}
\newcommand{\kr}{\,)}
\newcommand{\vvr}{(\vec{r})}
\newcommand{\vvrt}{(\vec{r},t)}
\newcommand{\vvrs}{(\vec{r}\,')}
\newcommand{\vint}{\int {\rm d}^3\vec{r}}
\newcommand{\vints}{\int {\rm d}^3\vec{r}\,'}
\newcommand{\pmu}{\partial_{\mu}}
\newcommand{\pnu}{\partial_{\nu}}
\newcommand{\pju}{\partial_{j}}

\newcommand{\pmo}{\partial^{\mu}}

\newcommand{\nmo}{\nabla^{\mu}}

\newcommand{\nmu}{\nabla_{\mu}}
\newcommand{\nnu}{\nabla_{\nu}}
\newcommand{\tr}{\text{tr}}
\newcommand{\nv}{\vec{\nabla}}
\newcommand{\nvs}{\vec{\nabla}'}
\newcommand{\brrs}{| \vec{r} - \vec{r}\,'|}
\newcommand{\eAm}{{}^{(1)}\!A^{\mu}(\vec{r})}
\newcommand{\eAn}{{}^{(1)}\!A^{\nu}(\vec{r})}
\newcommand{\eAo}{{}^{(1)}\!A_{0}(\vec{r})}
\newcommand{\eAou}{{}^{(1)}\!A_{0}(\vec{r})}
\newcommand{\zAou}{{}^{(2)}\!A_{0}(\vec{r})}
\newcommand{\aAo}{{}^{(a)}\!A_{0}(\vec{r})}
\newcommand{\Ava}{\vec{A}_a(\vec{r})}
\newcommand{\eAj}{{}^{(1)}\!A^{j}(\vec{r})}
\newcommand{\zAm}{{}^{(2)}\!A^{\mu}(\vec{r})}
\newcommand{\zAn}{{}^{(2)}\!A^{\nu}(\vec{r})}
\newcommand{\zAo}{{}^{(2)}\!A_{0}(\vec{r})}
\newcommand{\zAj}{{}^{(2)}\!A^{j}(\vec{r})}
\newcommand{\Ave}{\vec{A}_1(\vec{r})}
\newcommand{\Avz}{\vec{A}_2(\vec{r})}
\newcommand{\exAmu}{{}^{\textrm{(ex)}}\!A_{\mu}}
\newcommand{\exAou}{{}^{\textrm{(ex)}}\!A_{0}(\vec{r})}
\newcommand{\exAour}{{}^{\textrm{(ex)}}\!A_{0}(r)}
\newcommand{\exAju}{{}^{\textrm{(ex)}}\!A_{j}(\vec{r})}
\newcommand{\eAju}{{}^{(1)}\!A_{j}(\vec{r})}
\newcommand{\zAju}{{}^{(2)}\!A_{j}(\vec{r})}
\newcommand{\Amu}{A_{\mu}}
\newcommand{\Anu}{A_{\nu}}
\newcommand{\Aamu}{A^{a}_{\;\;\mu}}
\newcommand{\Aemu}{A^{1}_{\;\;\mu}}
\newcommand{\Aenu}{A^{1}_{\;\;\nu}}
\newcommand{\Azmu}{A^{2}_{\;\;\mu}}
\newcommand{\Aznu}{A^{2}_{\;\;\nu}}

\newcommand{\Aemo}{A^{1 \mu}}

\newcommand{\Azmo}{A^{2 \mu}}

\newcommand{\exAnu}{{}^{\textrm{(ex)}}\!A_{\nu}}
\newcommand{\aB}{a_{\rm B}}
\newcommand{\aM}{a_{\rm{M}}}

\newcommand{\Bm}{B^{\mu}(\vec{r})}
\newcommand{\Bn}{B^{\nu}(\vec{r})}
\newcommand{\Bo}{B^{0}(\vec{r})}
\newcommand{\Bou}{B_{0}(\vec{r})}
\newcommand{\Bj}{B^{j}(\vec{r})}
\newcommand{\Bju}{B_{j}(\vec{r})}
\newcommand{\Bms}{B^{\mu\,*}(\vec{r})}
\newcommand{\Bjus}{B_{j}^{*}(\vec{r})}
\newcommand{\Bos}{B^{0\,*}(\vec{r})}
\newcommand{\Bous}{B_{0}^{*}(\vec{r})}
\newcommand{\Bjs}{B^{j\,*}(\vec{r})}
\newcommand{\Bv}{\vec{B}(\vec{r})}
\newcommand{\Bvs}{\vec{B}^{*}(\vec{r})}
\newcommand{\Bmu}{B_{\mu}}
\newcommand{\Bmo}{B^{\mu}}
\newcommand{\Bnu}{B_{\nu}}

\newcommand{\Bsmu}{B_{\mu}^{*}}
\newcommand{\Bsnu}{B_{\nu}^{*}}
\newcommand{\Bsmo}{B^{* \mu}}

\newcommand{\B}{\vec{B}(\vec{r})}

\newcommand{\MDmu}{{\mathcal D}_{\mu}}
\newcommand{\MDmo}{{\mathcal D}^{\mu}}
\newcommand{\MDnu}{{\mathcal D}_{\nu}}
\newcommand{\MDno}{{\mathcal D}^{\nu}}
\newcommand{\MDou}{{\mathcal D}_0}

\newcommand{\MDju}{{\mathcal D}_j}

\newcommand{\MBDju}{{\mathbb D}_j}

\newcommand{\E}{\vec{E}}
\newcommand{\Ea}{\vec{E}_a(\vec{r})}
\newcommand{\Ee}{\vec{E}_1(\vec{r})}
\newcommand{\Ez}{\vec{E}_2(\vec{r})}
\newcommand{\aEj}{{}^{(a)}\!E^j(\vec{r})}
\newcommand{\Eva}{\vec{E}_a(\vec{r})}
\newcommand{\Ees}{E_{\textrm{es}}}
\newcommand{\EG}{E_{\textrm{G}}}
\newcommand{\EC}{E_{\textrm{C}}}
\newcommand{\ED}{E_{\textrm{D}}}
\newcommand{\ES}{E_{\textrm{S}}}
\newcommand{\ER}{E_{\textrm{R}}}
\newcommand{\ET}{E_{\textrm{T}}}
\newcommand{\EL}{E_{\textrm{L}}}
\newcommand{\Eint}{E_{\textrm{int}}}
\newcommand{\oEes}{\overset{\circ}{E}_{\textrm{es}}}
\newcommand{\Eees}{E^{(e)}_{\textrm{es}}}
\newcommand{\Emes}{E^{(m)}_{\textrm{es}}}
\newcommand{\ECh}{E_C^{\textrm{(h)}}}
\newcommand{\EGeh}{E_G^{\textrm{(eh)}}}
\newcommand{\ERe}{E_R^{\textrm{(e)}}}
\newcommand{\oERe}{\overset{\circ}{E}\!{}^{\,(e)}_R}
\newcommand{\ERm}{E_R^{\textrm{(m)}}}
\newcommand{\ECx}{E_C^{\textrm{(h)}}}
\newcommand{\ECg}{E_C^{\textrm{(g)}}}
\newcommand{\EGmg}{E_G^{\textrm{(mg)}}}
\newcommand{\ETmg}{E_T^{\textrm{(mg)}}}
\newcommand{\DETmg}{\Delta E_\textrm{T}^{\textrm{(mg)}}}
\newcommand{\tET}{\tilde{E}_\textrm{T}}
\newcommand{\ERSTe}{E_{\textrm{RST}}^{\textrm{(e)}}}
\newcommand{\DERSTe}{\Delta E_\textrm{RST}^{\textrm{(e)}}}
\newcommand{\Eexp}{E_\textrm{exp}}

\newcommand{\DERSTemg}{\Delta E_\textrm{RST}^{\textrm{(emg)}}}

\newcommand{\DERSTmg}{\Delta E_\textrm{RST}^{\textrm{(mg)}}}
\newcommand{\DEexp}{\Delta E_\textrm{exp}}
\newcommand{\DETx}{\Delta E_{\textrm{T}}^\textrm{(x)} }
\newcommand{\DETy}{\Delta E_{\textrm{T}}^\textrm{(y)} }
\newcommand{\DETz}{\Delta E_{\textrm{T}}^\textrm{(z)} }

\newcommand{\eFmnu}{{}^{(1)}\!F_{\mu \nu}(\vec{r})}
\newcommand{\zFmnu}{{}^{(2)}\!F_{\mu \nu}(\vec{r})}
\newcommand{\eFmno}{{}^{(1)}\!F^{\mu \nu}(\vec{r})}
\newcommand{\zFmno}{{}^{(2)}\!F^{\mu \nu}(\vec{r})}
\newcommand{\Fmnu}{F_{\mu \nu}}
\newcommand{\SMFmnu}{{}^{(S)}\!{\mathcal F}_{\mu \nu}}
\newcommand{\exMFmnu}{{}^{\textrm{(ex)}}\!{\mathcal F}_{\mu \nu}}
\newcommand{\exFmnu}{{}^{\textrm{(ex)}}\!F_{\mu \nu}}
\newcommand{\Fnulo}{F_{\nu}^{\;\;\lambda}}
\newcommand{\Fenulo}{F^{1\;\;\lambda}_{\;\;\nu}}
\newcommand{\Fznulo}{F^{2\;\;\lambda}_{\;\;\nu}}
\newcommand{\Fsolo}{F^{\sigma \lambda}}

\newcommand{\Fesulu}{F^{1}_{\;\;\sigma \lambda}}
\newcommand{\Fzsolo}{F^{2 \sigma \lambda}}

\newcommand{\exFsulu}{{}^{\textrm{(ex)}}\!F_{\sigma \lambda}}
\newcommand{\exFnulo}{{}^{\textrm{(ex)}}\!F_{\nu}^{\;\;\lambda}}
\newcommand{\Famnu}{F^{a}_{\;\;\mu \nu}}
\newcommand{\Femnu}{F^{1}_{\;\;\mu \nu}}
\newcommand{\Femlu}{F^{1}_{\;\;\mu \lambda}}
\newcommand{\Fzmnu}{F^{2}_{\;\;\mu \nu}}
\newcommand{\Fzmlu}{F^{2}_{\;\;\mu \lambda}}
\newcommand{\MFmnu}{{\mathcal F}_{\mu \nu}}
\newcommand{\xefnu}{{}^{\textrm{(xe)}}\!f_{\nu}}
\newcommand{\fs}{f_{*}}

\newcommand{\Gmnu}{G_{\mu \nu}(\vec{r})}

\newcommand{\Gmnus}{G^{*}_{\mu \nu}(\vec{r})}

\newcommand{\Gmnos}{G^{* \mu \nu \,}(\vec{r})}
\newcommand{\Gnulo}{G_{\nu}^{\;\;\lambda}}
\newcommand{\Gmn}{G_{\mu \nu}}
\newcommand{\Gml}{G_{\mu \lambda}}
\newcommand{\Glso}{G^{\sigma \lambda}}
\newcommand{\Gsnulo}{G^{*\lambda}_{\nu}}
\newcommand{\Gsmnu}{G^{*}_{\mu \nu}}
\newcommand{\Gsslu}{G^{*}_{\sigma \lambda}}
\newcommand{\Gsmlu}{G^{*}_{\mu \lambda}}

\newcommand{\hm}{h_{\mu}(\vec{r})}
\newcommand{\ho}{h_{0}(\vec{r})}
\newcommand{\hj}{h_{j}(\vec{r})}
\newcommand{\hms}{h_{\mu}^{*}(\vec{r})}
\newcommand{\hos}{h_{0}^{*}(\vec{r})}
\newcommand{\hoss}{h_{0}^{*}(\vec{r}\,')}
\newcommand{\hjs}{h_{j}^{*}(\vec{r})}
\newcommand{\hv}{\vec{h}(\vec{r})}

\newcommand{\hvs}{\vec{h}^{*}(\vec{r})}
\newcommand{\hvsp}{\vec{h}^{*}(\vec{r}\,')}
\newcommand{\hmu}{h_{\mu}}
\newcommand{\hnu}{h_{\nu}}

\newcommand{\hsmu}{h^{*}_{\mu}}
\newcommand{\hsnu}{h^{*}_{\nu}}

\newcommand{\Hv}{\vec{H}}
\newcommand{\Ha}{\vec{H}_a(\vec{r})}
\newcommand{\He}{\vec{H}_1(\vec{r})}
\newcommand{\Hz}{\vec{H}_2(\vec{r})}
\newcommand{\aHj}{{}^{(a)}\!H^j(\vec{r})}
\newcommand{\Hva}{\vec{H}_a(\vec{r})}
\newcommand{\Hex}{\vec{H}_{ex}}
\newcommand{\MHmo}{{\mathcal H}^{\mu}}
\newcommand{\MHmu}{{\mathcal H}_{\mu}}
\newcommand{\MHbmu}{\bar{{\mathcal H}}_{\mu}}

\newcommand{\MHnu}{{\mathcal H}_{\nu}}

\newcommand{\MI}{{\mathcal I}}

\newcommand{\Djm}{{}^{(D)}\!j_{\mu}}

\newcommand{\jmu}{j_{\mu}}
\newcommand{\exjnu}{{}^{\textrm{(ex)}}\!j_{\nu}}
\newcommand{\exjmu}{{}^{\textrm{(ex)}}\!j_{\mu}}
\newcommand{\exjmo}{{}^{\textrm{(ex)}}\!j^{\mu}}
\newcommand{\MJmu}{{\mathcal J}_{\mu}}
\newcommand{\MJnu}{{\mathcal J}_{\nu}}
\newcommand{\exMJmu}{{}^{\textrm{(ex)}}\!{\mathcal J}_{\mu}}
\newcommand{\exMJnu}{{}^{\textrm{(ex)}}\!{\mathcal J}_{\nu}}
\newcommand{\SMJmu}{{}^{\textrm{(S)}}\!{\mathcal J}_{\mu}}
\newcommand{\SMJnu}{{}^{\textrm{(S)}}\!{\mathcal J}_{\nu}}

\newcommand{\ekm}{{}^{(1)}\!k_{\mu}(\vec{r})}
\newcommand{\eko}{{}^{(1)}\!k_{0}(\vec{r})}
\newcommand{\akm}{{}^{(a)}\!k_{\mu}(\vec{r})}
\newcommand{\akmos}{{}^{(a)}\!k^{\mu}(\vec{r}\,')}
\newcommand{\ako}{{}^{(a)}\!k_{0}(\vec{r})}
\newcommand{\akos}{{}^{(a)}\!k_{0}(\vec{r}\,')}
\newcommand{\ekj}{{}^{(1)}\!k_{j}(\vec{r})}
\newcommand{\zkm}{{}^{(2)}\!k_{\mu}(\vec{r})}
\newcommand{\zko}{{}^{(2)}\!k_{0}(\vec{r})}
\newcommand{\zkop}{{}^{(2)}\!k_{0}(\vec{r}\,')}
\newcommand{\zkj}{{}^{(2)}\!k_{j}(\vec{r})}
\newcommand{\kv}{\vec{k}(\vec{r})}
\newcommand{\kvs}{\vec{k}(\vec{r}\,')}
\newcommand{\kva}{\vec{k}_a(\vec{r})}
\newcommand{\kvas}{\vec{k}_a(\vec{r}\,')}
\newcommand{\kve}{\vec{k}_1(\vec{r})}

\newcommand{\kvz}{\vec{k}_2(\vec{r})}
\newcommand{\kvzs}{\vec{k}_2(\vec{r}\,')}
\newcommand{\dkm}{{}^{(d)}\!k_{\mu}}
\newcommand{\pkm}{{}^{(p)}\!k_{\mu}}
\newcommand{\kmor}{k^{\mu}(\vec{r})}
\newcommand{\kmu}{k_{\mu}}
\newcommand{\kamu}{k^{a}_{\;\;\mu}}
\newcommand{\kemu}{k^{1}_{\;\;\mu}}
\newcommand{\kzmu}{k^{2}_{\;\;\mu}}

\newcommand{\kenu}{k^{1}_{\;\;\nu}}
\newcommand{\kznu}{k^{2}_{\;\;\nu}}

\newcommand{\kdr}{\vec{k}_d(\vec{r})}
\newcommand{\kpr}{\vec{k}_p(\vec{r})}

\newcommand{\Maem}{M_a^{(em)}}
\newcommand{\Maxg}{M_a^{(hg)}}
\newcommand{\Meem}{M_{1}^{(em)}}
\newcommand{\Mzem}{M_{2}^{(em)}}
\newcommand{\Mexg}{M_{1}^{(hg)}}
\newcommand{\Mzxg}{M_{2}^{(hg)}}
\newcommand{\Mae}{M_a^{(e)}}
\newcommand{\Mam}{M_a^{(m)}}
\newcommand{\Mee}{M_1^{(e)}}
\newcommand{\Mze}{M_2^{(e)}}
\newcommand{\Mem}{M_1^{(m)}}
\newcommand{\Mzm}{M_2^{(m)}}
\newcommand{\Max}{M_a^{(h)}}
\newcommand{\Mex}{M_1^{(h)}}
\newcommand{\Mzx}{M_2^{(h)}}
\newcommand{\Mag}{M_a^{(g)}}
\newcommand{\Mamg}{M_a^{(mg)}}

\newcommand{\Meg}{M_1^{(g)}}
\newcommand{\Mzg}{M_2^{(g)}}
\newcommand{\Maex}{M_a^{(eh)}}
\newcommand{\Ms}{M_{*}}

\newcommand{\tMo}{\tilde{M}_0}

\newcommand{\aRpm}{{}^{(a)}\!R_{\pm}(r)}

\newcommand{\eRp}{{}^{(1)}\!R_{+}(r)}
\newcommand{\eRm}{{}^{(1)}\!R_{-}(r)}

\newcommand{\zRp}{{}^{(2)}\!R_{+}(r)}
\newcommand{\zRm}{{}^{(2)}\!R_{-}(r)}
\newcommand{\Rp}{R_{+}(r)}
\newcommand{\oRp}{\overset{\circ}{R}_{+}(r)}
\newcommand{\Rm}{R_{-}(r)}
\newcommand{\Rpm}{R_{\pm}(r)}
\newcommand{\MRp}{{\mathbb R}_{+}}

\newcommand{\STmn}{{}^{\textrm{(S)}}\!T_{\mu \nu}}
\newcommand{\esTmn}{{}^{\textrm{(es)}}\!T_{\mu \nu}}
\newcommand{\exTmn}{{}^{\textrm{(ex)}}\!T_{\mu \nu}}
\newcommand{\as}{\alpha_{\rm{S}}}
\newcommand{\TTmnu}{{}^{\textrm{(T)}}\!T_{\mu \nu}}
\newcommand{\STmnu}{{}^{\textrm{(S)}}\!T_{\mu \nu}}
\newcommand{\esTmnu}{{}^{\textrm{(es)}}\!T_{\mu \nu}}
\newcommand{\DTmnu}{{}^{\textrm{(D)}}\!T_{\mu \nu}}
\newcommand{\GTmnu}{{}^{\textrm{(G)}}\!T_{\mu \nu}}
\newcommand{\CTmnu}{{}^{\textrm{(C)}}\!T_{\mu \nu}}
\newcommand{\RTmnu}{{}^{\textrm{(R)}}\!T_{\mu \nu}}
\newcommand{\TToo}{{}^{\textrm{(T)}}\!T_{00}}
\newcommand{\RToo}{{}^{\textrm{(R)}}\!T_{00}}

\newcommand{\GToo}{{}^{\textrm{(G)}}\!T_{00}}
\newcommand{\CToo}{{}^{\textrm{(C)}}\!T_{00}}
\newcommand{\esToo}{{}^{\textrm{(es)}}\!T_{00}}
\newcommand{\DToo}{{}^{\textrm{(D)}}\!T_{00}}

\newcommand{\Wp}{\vec{W}_p(\vartheta, \varphi)}
\newcommand{\Wps}{\vec{W}^{*}_p(\vartheta, \varphi)}

\newcommand{\X}{\vec{X}(\vec{r})}
\newcommand{\Xs}{\vec{X}^{*}(\vec{r})}

\newcommand{\Y}{\vec{Y}(\vec{r})}
\newcommand{\Ys}{\vec{Y}^{*}(\vec{r})}

\newcommand{\zex}{z_{\textrm{ex}}}

\newcommand{\es}{\varepsilon_{*}}

\newcommand{\zeoe}{\zeta^{\frac{1}{2},\frac{1}{2}}_{\;\;\:0}}
\newcommand{\zeome}{\zeta^{\frac{1}{2},-\frac{1}{2}}_{\;\;\:0}}
\newcommand{\zeee}{\zeta^{\frac{1}{2},\frac{1}{2}}_{\;\;\:1}}
\newcommand{\zeeme}{\zeta^{\frac{1}{2},-\frac{1}{2}}_{\;\;\:1}}
\newcommand{\zjlm}{\zeta^{j,m}_{\;\;l}}

\newcommand{\vmL}{\vec{\mu}_{{\rm L}}}
\newcommand{\vmS}{\vec{\mu}_{{\rm S}}}
\newcommand{\vmJ}{\vec{\mu}_{{\rm J}}}
\newcommand{\mB}{\mu_{{\rm B}}}

\newcommand{\Smno}{\Sigma^{\mu \nu}}

\newcommand{\bp}{\bar{\psi}}
\newcommand{\bpr}{\bar{\psi}(\vec{r})}
\newcommand{\pr}{\psi(\vec{r})}

\newcommand{\afp}{{}^{(a)}\!\phi_{+}}
\newcommand{\efp}{{}^{(1)}\!\phi_{+}}
\newcommand{\zfp}{{}^{(2)}\!\phi_{+}}
\newcommand{\afm}{{}^{(a)}\!\phi_{-}}
\newcommand{\efm}{{}^{(1)}\!\phi_{-}}
\newcommand{\zfm}{{}^{(2)}\!\phi_{-}}

\begin{document}

\setcounter{page}{0}
\title{\bf Magnetic Interactions \\in\\Relativistic Two-Particle Systems}
\author{P.\ Schust and M.\ Sorg\\ II.\ Institut f\"ur Theoretische Physik der
Universit\"at Stuttgart\\ Pfaffenwaldring 57 \\ D 70550 Stuttgart, Germany \vspace{6cm}}
\begin{abstract}
  The magnetic interactions of the two electrons in helium-like ions are studied in detail
  within the framework of Relativistic Schr\"odinger Theory (RST). The general results are
  used to compute the ground-state interaction energy of some highly-ionized atoms ranging
  from germanium~$(Z=32)$ up to bismuth~$(Z=83)$. When the \emph{magnetic} interaction
  energy is added to its electric counterpart resulting from the electrostatic
  approximation, the present RST predictions reach a similar degree of precision (relative
  to the experimental data) as the other theoretical approaches known in the literature.
  However since the RST magnetism is then treated only in lowest-order approximation,
  further improvements of the RST predictions seem possible.
\end{abstract}
\maketitle

\section{Introduction and Survey of Results}

The overwhelming success of relativistic quantum field theory (especially quantum
electrodynamics) could tempt one to belief that any relativistic form of quantum theory
must necessarily include the virtual degrees of freedom of matter. Indeed it is well-known
that the virtual processes show up experimentally in form of radiation corrections, vacuum
polarization, pair creation and annihilation etc.~\cite{We}. If such a viewpoint would be
true, it would logically not be possible to construct a consistent relativistic
\emph{quantum mechanics} for interacting~$N-$particle systems whose non relativistic
description presents no problem at all~\cite{Ma}. However it seems that there actually
exists no convincing argument why such a \emph{relativistic} quantum mechanics for systems
of \emph{fixed particle number}~$N$ could not exist. Quite on the contrary, since the
virtual processes of relativistic field theory (e.g.\ the Lamb shift) induce only small
corrections of the results obtained from the first-quantized quantum mechanics theories,
one concludes that a consistent relativistic quantum mechanics would correctly describe
the main relativistic effects occurring in the many-particle systems (e.g.\ heavy atoms)
but would leave the small corrections due to the virtual processes to the impact of
quantum field theory.

In this context, one could now think that the desired relativistic quantum mechanics
should be deduced as a certain kind of limit process from quantum field theory itself,
namely through fixing the particle number by neglection of just the virtual
processes. Such a deduction has been tried already long ago by Bethe and
Salpeter~\cite{Sa,Ge} but the resulting~$N-$particle wave equations are plagued by many
deficiencies and interpretation problems~\cite{G,La}. Therefore a new approach to
relativistic quantum mechanics for many-particle systems has recently been established in
form of Relativistic Schr\"odinger Theory (RST)~\cite{So,RuSo,Ru,Sch}. This approach is a
relativistic gauge theory based upon the gauge group~U(N). The gauge group is broken
down to the abelian subgroup~U(1)$\times$U(1)\ldots$\times$U(1) which then describes the
electromagnetic interaction of~$N$ particles, whereas the frozen gauge degrees of freedom
are responsible for the exchange interactions. The non-relativistic limit of this theory
coincides with the well-known Hartree-Fock approach~\cite{Ve}.

The present paper presents a test of the practical usefulness of RST, namely by
elaborating its predictions for the ground-state interaction energy of the two electrons
in highly-ionized helium-like atoms where relativistic effects play a non-negligible role.
In contrast to the two-particle ground-state energy itself, the corresponding
\emph{interaction} energy is directly accessible to the experiments~\cite{MES} which thus
provide sufficient data for testing various theoretical approaches: relativistic~$1/Z$
expansion~\cite{Dr} multiconfiguration, Dirac-Fock method (MCDF)~\cite{F1,F2,Go},
relativistic many-body perturbation theory (MBPT)~\cite{Wi,Jo}, all-order technique for
relativistic MBPT~\cite{Pl}. These four theoretical approaches have been chosen to be
opposed to the experimental data; and the present paper adds now the corresponding RST
results which are found to describe the experimental data with a similar degree of
precision as these other theoretical approaches (table~II). This success is attained by
going beyond the electrostatic approximation~(\cite{Ve2}) and taking into account also the
\emph{magnetic} interactions which implies an improvement of the RST predictions by
(roughly) one order of magnitude. However one should observe here that these RST results
have been obtained in a preliminary way by a very rough approximation technique, namely by
neglecting the non-abelian character of RST through linearizing the gauge field equations
(Sect.~4B) and furthermore by treating the magnetic interactions only in their
non-relativistic limit. Thus the potentiality of RST is not yet fully exploited and
further improvements of the RST predictions seem still possible.

The procedure is as follows: In Sect.~II we present a brief survey of RST in order that
the main arguments of the subsequent discussions can be understood without looking up the
whole development of the theory in the preceding papers. Here the emphasis is laid upon
the construction of an energy functional~$\ET$, namely as the integral (over
all 3-space) of an energy density~$T_{00}(\vec{r})$, see equation~(\ref{244})
below. Clearly the reason is that, with such a functional at hand, the energy level system
of the bound~$N-$particle systems may be determined as the value of that
functional~$E_{\mathrm{T}}$ upon the stationary bound solutions of the RST field
equations.

This kind of solutions is then described in great detail in Sect.~III with an explicit
presentation of the mass-eigenvalue equations (see equations~(\ref{34a})-(\ref{34b})
below). Since this eigenvalue problem is too difficult to be solved exactly, one has to
resort to appropriate approximation techniques (as is mostly the case in quantum field
theory). Here the logical structure of RST suggests to first neglect the \emph{magnetic}
interactions between the particles (\emph{``electrostatic approximation''}) which leaves us
with a simpler eigenvalue problem (see equations~(\ref{36a})-(\ref{36b}) below). This
truncated system is then solved numerically for the ground state of the two-particle
system in the Coulomb field~$\exAou$~(\ref{39}) where the nuclear charge numbers~$(\zex)$
range from~$\zex=32$ (germanium) up to~$\zex=83$ (bismuth). The corresponding ground-state
interaction energy~$\DERSTe$ (in the electrostatic approximation) is then compared to the
corresponding experimental values~$\DEexp$ (see table~I); and it is found that there is a
discrepancy between the theoretical (RST) and experimental values extending from~1.7~eV
for germanium up to 11.5~eV for bismuth. However some intuitive arguments indicate that
the observed discrepancy should actually be due to the neglection of the magnetic
forces. When the latter forces are taken into account, there arises a structure equation
which specifies the desired interaction energy~$\Delta E$ in terms of the electromagnetic
coupling constant~$(\zex\as)$ and two functions~$\es$ and~$\fs^2$ which are only weakly
depending upon that coupling constant, see equation~(\ref{358n}) and the last two columns
of table~I.

In Sect.~IV the hypothesis of the magnetic origin of the discrepancy between the
electrostatic RST predictions~$\DERSTe$ and the experimental data~$\DEexp$ is inspected
very thoroughly by working out the RST field theory of atomic magnetism. Here it seems
reasonable to consider first the magnetic effects in the lowest-order approximation, i.e.\
we linearize the non-abelian gauge field equations and additionally we resort to the
non-relativistic limit of the electrostatic wave functions for calculating the magnetic
energy contributions. It is very striking to observe that the magnetic and electric
contributions display certain dissimilarities which can however be revealed as necessary
consequences of the principle of minimal coupling and Lorentz invariance of the
theory. Though one obtains a very plausible result for the magnetic energy
contribution~$\DETmg$, see equation~(\ref{457}) below, one nevertheless may wish to
become convinced by an independent argument which supports our claim that magnetism is
correctly incorporated RST by the present approach.

And indeed, such an additional argument in favor of the RST picture of atomic magnetism
can be supplied, namely by considering the interaction of the bound particles with an
\emph{external} magnetic field~$\Hex$ (Sect.~V). Here it can be shown that the interaction
energy between the particles and the external magnetic field~$\Hex$ exactly agrees with
the conventional results for the Zeeman effects; i.e.\ the magnetic RST energy coincides
with the expectation value of the conventional Zeeman Hamiltonian, see
equation~(\ref{526}) below.

Being thus convinced of the physical correctness of the RST picture of atomic magnetism,
one can return to the hypothesis of the magnetic origin of the numerical gap between the
electrostatic RST predictions~$\DERSTe$ and the experimental data~$\DEexp$ of table~I. In
Sect.~VI we apply the general results of Sect.~IV in order to once more calculate the
ground-state interaction energy for the helium-like ions from~$\zex=32$ (germanium) up
to~$\zex=83$ (bismuth), but now with inclusion of the magnetic interactions. Here we are
satisfied for the moment with their \emph{linearized} description in the
\emph{non-relativistic} limit. Amazingly enough, this simple treatment of atomic magnetism
is sufficient in order to close the gap between the RST predictions~$\DERSTe$ and the
experimental data~$\DEexp$ up to less than 0.5\%, see table~II in comparison to table~I.
This must be considered as a rather convincing argument in favor of the ``magnetic''
hypothesis for the observed discrepancy between~$\DERSTe$ and~$\DEexp$. It is also very
satisfying to observe that, with the inclusion of the magnetic effects, the corresponding
RST predictions~$\DERSTemg$ are now as close to the experimental values~$\DEexp$ as is the
case with the other approximation methods known in the literature: Relativistic Many-Body
Perturbation Theory~\cite{Wi,Jo}, all-order technique for MBPT~\cite{Pl},
Multi-Configuration Dirac-Fock method (MCDF)~\cite{F1,F2,Go} and relativistic 1/Z
expansion~\cite{Dr}, see table~II. (The predictions of these theoretical approaches have
been listed in table~III of ref.~\cite{MES}. Clearly one expects that the RST predictions
will be further improved by taking into account also the \emph{relativistic} effects of
atomic magnetism and retaining the non-linear terms due to the non-abelian character of
the gauge group~U(N) (separate paper).

Finally it should be stressed that the numerical success of the magnetic hypothesis is
mainly due to the application of the non-abelian~U(2). The reason is that after the
\emph{``abelian symmetry-breaking''} (Sect.~2) the frozen gauge degrees of freedom of~U(2)
imply the existence of a ``magnetic'' exchange vector potential~$\B$ which plays an
analogous part with respect to the x- and y-axis as the magnetostatic potentials~$\Ava$
with respect to the z-axis of the coordinate system. Indeed when the spins of the two
ground-state electrons are oriented along the z-axis, the corresponding spin-spin
interaction energy~$\DETz$ (see equation~(\ref{571}) below) is well suited in order to
contribute to closing the gap between~$\DERSTe$ and~$\DEexp$, however only one third of
the gap could be closed in this way. The other two thirds of the gap must be filled by the
spin-spin interaction energies~$\DETx$~(\ref{573a}) and~$\DETy$~(\ref{573b}) due to the
spin orientation along the x- and y-axis (see fig.~1). But the latter interactions are
described by just that exchange vector potential~$\B$ being due to the frozen gauge
degrees of freedom of~U(2). Thus each axis of spin orientation contributes the same
magnetic energy, and this is nothing else than a demonstration of the \emph{isotropic}
geometry of the ground-state. In this sense the choice of the non-abelian~U(2) is seen
to be necessary just in order to guarantee the isotropy of the ground state.



\section{Relativistic Schr\"odinger Theory}
In order to let the subsequent elaboration of arguments appear sufficiently self\bs contained, we first present a brief sketch of the {\bf R}elativistic {\bf S}chr\"odinger {\bf T}heory \kl RST\kr, which places emphasis on a closer inspection of the energy functional, since this will ultimately yield the energy level system of the bound many\bs particle systems. For more details and deductions, the interested reader is referred to the preceding papers, e.g. refs.~(\cite{So})-(\cite{Ve}).

\subsection{RST Dynamics}

As for any field theory of matter, the fundamentals of RST consist in a basic system of field equations, which is subdivided into three coupled subystems: matter dynamics, Hamiltonian dynamics and gauge field dynamics. \\
\noindent {\large\it (i) Matter Dynamics}\ \\
\indent When matter occurs in form of pure state $\Psi$, its distribution over space\bs time is governed by the {\bf R}elativistic {\bf S}chr\"odinger {\bf E}quation \kl RSE\kr
\begin{equation}
\label{21}
i \hbar c \MDmu \Psi = \MHmu \Psi \;.
\end{equation}
\noindent If it is more adequate to describe matter by a mixture, being characterized by an intensity matrix $\MI$, one requires $\MI$ to obey the {\bf R}elativistic von {\bf N}eumann {\bf E}quation \kl RNE\kr
\begin{equation}
\label{22}
\MDmu \MI =  \frac{i}{\hbar c} \left[ \MI \cdot \MHbmu - \MHmu \cdot \MI \right] \;.
\end{equation}
 \noindent Clearly, a pure state $\Psi$ can be considered as a special type of mixture, namely that for which the intensity matrix degenerates to the tensor product of $\Psi$ and its Hermitian conjugate $\bar{\Psi}$, i.e.
\begin{equation}
\label{23}
\MI \Rightarrow \Psi \otimes \bar{\Psi} \;.
\end{equation}
\noindent In the present paper we are restricting ourselves to the investigation of stationary bound systems being described by pure states. Since a pure state $\Psi$ of an $N$\bs fermion system has $4N$ components, we have to consider a ${\mathbb C}^{4N}$\bs realization of RST. \\
\noindent{\large\it (ii) Hamiltonian Dynamics} \\
\indent The Hamiltonian $\MHmu$, occuring in the matter field equations \rf{21} and \rf{22}, is itself a dynamical object obeying its own field equations, namely the {\it integrability condition}
\begin{equation}
\label{24}
\MDmu \MHnu - \MDnu \MHmu + \frac{i}{\hbar c} \left[ \MHmu, \MHnu \right] = i\hbar c \MFmnu
\end{equation}
\noindent and the {\it conservation equation} which reads for fermions
\begin{equation}
\label{25}
\MDmo \MHmu - \frac{i}{\hbar c} \MHmo \cdot \MHmu = i \hbar c \left(\frac{M c}{\hbar} \right)^2 \cdot {\bf 1} - i \hbar c \Smno \MFmnu \;.
\end{equation}
\noindent For bosons the spin term \kl last term on the right\bs hand side\kr\;is omitted. The meaning of the integrability condition \rf{24} is to ensure the validity of the bundle identities
\begin{subequations}
\begin{align}
\label{26a}
\left[ \MDmu\MDnu - \MDnu\MDmu\right] \Psi &= \MFmnu \cdot \Psi \\
\label{26b}
\left[ \MDmu\MDnu - \MDnu\MDmu\right] \MI &= \left[\MFmnu, \MI \right] \;.
\end{align}
\end{subequations}
\noindent Similarly, the conservation equation \rf{25} guarantees the existence of certain conservation laws, e.g. the charge conservation
\begin{equation}
\label{27}
\nmo \jmu \equiv 0 \;,
\end{equation}
\noindent or the energy\bs momentum conservation for free particles \kl Dirac particles here\kr
\begin{equation}
\label{28}
\nmo \STmn = 0 \;,
\end{equation}
\noindent whose energy-momentum density is denoted by $\STmn$.\\
\indent Clearly, if the considered $N$\bs particle system is interacting with an external field $\exMFmnu$
\begin{equation}
\label{29}
\exMFmnu = -i\, \exFmnu \cdot {\bf 1} \equiv \MFmnu - \SMFmnu \;,
\end{equation}
\noindent the energy\bs momentum density $\STmn$ cannot obey the conservation law \rf{28} without being complemented by the energy\bs momentum density $\esTmn$ due to the external interaction. Consequently, the conservation law \rf{28} has to be replaced by the balance equation
\begin{equation}
\label{210}
\nmo\left(\STmn + \esTmn \right) = - \xefnu \;,
\end{equation}
\noindent where the Lorentz\bs force density $\xefnu$ is given by
\begin{equation}
\label{211}
\xefnu = -\hbar c \Fmnu \exjmo \;.
\end{equation}
\noindent Here, $\exjmu$ is the external four\bs current generating the external field $\exFmnu$ according to Maxwell's equations
\begin{eqnarray}
\label{212}
\nmo \exFmnu &=& 4\pi \as \exjnu \\
\nonumber
\Bigg(\as &=& \frac{e^2}{\hbar c}\Bigg) \;,
\end{eqnarray}
\noindent and $\Fmnu$ is the coherent electromagnetic field strength generated by all the particles of the system:
\begin{equation}
\label{213}
\Fmnu = \frac{i}{N-1} \text{tr}\left\{\SMFmnu\right\} \;.
\end{equation}
\noindent This means that the source of this total field $\Fmnu$ is the total current $\jmu$ of the $N$\bs particle system
\begin{equation}
\label{214}
\nmo \Fmnu = -4\pi\as j_{\nu} \;,
\end{equation}
\noindent which obeys the strict conservation law \rf{27}.If one prefers to think of a closed system, one adds the energy\bs momentum density $\exTmn$ of the external source
\begin{equation}
\label{215}
\nmo \exTmn = \xefnu
\end{equation}
\noindent and then finds the desired closedness relation by combining equations \rf{210} and \rf{215}
\begin{equation}
\label{216}
\nmo \left( \STmn + \esTmn + \exTmn \right) = 0 \;.
\end{equation}
\indent Finally, let us also remark that the conservation equation \rf{25} is equivalent to the following condition upon the Hamiltonian:
\begin{equation}
\label{217}
\Gamma^{\mu} \cdot \MHmu = Mc^2 {\bf 1} \;.
\end{equation}
\noindent Here, the total velocity operator $\Gamma^{\mu}$ generates a $(4N)$\bs dimensional representation of the Clifford algebra $C(1,3)$, i.e.
\begin{equation}
\label{218}
\Gamma_{\mu} \cdot \Gamma_{\nu} + \Gamma_{\nu} \cdot \Gamma_{\mu} = 2 g_{\mu \nu} \cdot {\bf 1} \;,
\end{equation}
\noindent with the corresponding generators $\Sigma_{\mu \nu}$ of the group $Spin(1,3)$ being given by
\begin{equation}
\label{219}
\Sigma_{\mu \nu} = \frac{1}{4} \left[ \Gamma_{\mu}, \Gamma_{\nu} \right]\ ,
\end{equation}
see the right-hand side of the conservation equation~\ref{25}. The physical meaning of
$\Gamma_{\mu}$ is to build up the {\it total current} $\jmu$ according to
\begin{equation}
\label{220}
\jmu \doteqdot \text{tr} \left\{ \MI \cdot \Gamma_{\mu} \right\} \;,
\end{equation}
\noindent i.e. for a pure state \rf{23}
\begin{equation}
\label{221}
\jmu = \bar{\Psi} \cdot \Gamma_{\mu} \cdot \Psi \;.
\end{equation}
\noindent Since by means of the latter form \rf{217} of the conservation equation the RSE \rf{21} can be converted to the $N$\bs particle Dirac equation
\begin{equation}
\label{222}
i\hbar \Gamma^{\mu} \MDmu \Psi = Mc \Psi \;,
\end{equation}
\noindent it thus becomes a simple matter to verify directly the charge conservation law \rf{27}. This form of the charge conservation does not only work for pure states but equally well for mixtures, which can be proven by resorting to the RNE \rf{22} instead of the RSE \rf{21}. Observe also that the non\bs relativistic limit of the $N$\bs particle Dirac equation \rf{222} coincides with the well\bs known Hartree\bs Fock equations \cite{Ve}.\\
\noindent (iii) {\large\it Gauge Field Dynamics} \\
\indent In order to have a closed dynamical system, one finally has to specify a field equation for the bundle curvature \kl{\it ``field strength''}\kr\;$\MFmnu$. This is the non\bs abelian Maxwell equation
\begin{equation}
\label{223}
\MDmo \MFmnu = -4\pi i\as \MJnu \;.
\end{equation}
\noindent Similarly, as it was done for the field strenght $\MFmnu$ \rf{29}, the current operator $\MJmu$ may also be split into an external and a system part
\begin{subequations}
\begin{align}
\label{224a}
\MJmu &= \exMJmu + \SMJmu \\
\label{224b}
\exMJmu &= \exjmu \cdot {\bf 1}
\end{align}
\end{subequations}
\noindent so that the non\bs abelian Maxwell equation \rf{223} can be required to decay analogously into two parts
\begin{subequations}
\begin{align}
\label{225a}
\MDmo \exMFmnu &= -4\pi i\as \exMJnu \\
\label{225b}
\MDmo \SMFmnu &= -4\pi i\as \SMJnu \;.
\end{align}
\end{subequations}
\noindent Here the external part \rf{225a} gives the former \kl abelian\kr\;Maxwell equation \rf{212}, whereas the ``total'' Maxwell equation \rf{214} emerges as the trace part of \rf{225b} with
\begin{equation}
\label{226}
\jmu = - \frac{1}{N-1} \tr\left\{\MJmu\right\} \;.
\end{equation}
\indent The essential point with the gauge structure of RST refers to the right choice of
the gauge group. In conventional quantum \kl field\kr\;theory, the group $U(1)$ is evoked
in order to describe the electromagnetic interactions. However, in RST the $N$\bs particle
bundle of wave functions $\Psi$ is the Whitney sum of the corresponding single\bs particle
bundles which suggests to adopt the $N$\bs fold product group $U(1)\times
U(1)\times\ldots\times U(1)$ as the adequate subgroup of $U(N)$ for the description of the
RST gauge interactions. Though the {\it electromagnetic} interactions
are thus adequately incorporated into RST by means of this \emph{``abelian symmetry breaking''}, one needs the residual structure of the embedding group $U(N)$ in order to take into account also the {\it exchange} interactions between the $N$ particles. \\
\indent Considering for instance the two\bs fermion case \kl$N=2$\kr, one has to deal with two ``electromagnetic generators'' $\tau_a \; (a=1,2)$ and two ``exchange generators'' $\chi,\bar{\chi}$ in order to span the gauge algebra ${\mathfrak u}(2)$. The bundle connection \kl``gauge potential''\kr\;${\mathcal A}_{\mu}$ may then again be splitted up into an external $(ex)$ and internal system part $(S)$
\begin{subequations}
\begin{align}
\label{227a}
{\mathcal A}_{\mu} &= {}^{(ex)}\!{\mathcal A}_{\mu} + {}^{(S)}\!{\mathcal A}_{\mu} \\
\label{227b}
{}^{(ex)}\!{\mathcal A}_{\mu} &= -i \exAmu \cdot{\bf 1} \;,
\end{align}
\end{subequations}
\noindent where the internal part ${}^{(S)}\!{\mathcal A}_{\mu}$ decomposes with respect to the ${\mathfrak u}(2)$ basis $\left\{ \tau_a,\chi,\bar{\chi} \right\}$ as follows
\begin{equation}
\label{228}
{}^{(S)}\!{\mathcal A}_{\mu} = \Aamu \tau_a + \Bmu \chi - \Bsmu \bar{\chi} \;.
\end{equation}
\noindent Clearly, the connection ${\mathcal A}_{\mu}$ generates its curvature $\MFmnu$ as usual
\begin{equation}
\label{229}
\MFmnu = \nmu {\mathcal A}_{\nu} - \nnu {\mathcal A}_{\mu} + \left[{\mathcal A}_{\mu},{\mathcal A}_{\nu}\right] \;,
\end{equation}
\noindent and by decomposing the internal part $\SMFmnu$ \rf{29} in a similar way
\begin{equation}
\label{230}
\SMFmnu = \Famnu \tau_a + \Gmn \chi - \Gsmnu \bar{\chi} \;,
\end{equation}
\noindent the curvature components $\Famnu, \Gmn$ are found in terms of the connection components $\Aamu,\Bmu$ as follows:
\begin{subequations}
\begin{align}
\label{231a}
\exFmnu &= \nmu \exAnu - \nnu \exAmu \\
\label{231b}
\Femnu &= \nmu \Aenu - \nnu \Aemu + i \left[\Bmu\Bsnu - \Bnu\Bsmu\right] \\
\label{231c}
\Fzmnu &= \nmu \Aznu - \nnu \Azmu - i \left[\Bmu\Bsnu - \Bnu\Bsmu\right] \\
\label{231d}
\Gmn &= \nmu \Bnu - \nnu \Bmu + i\left[\Aemu - \Azmu \right]\cdot\Bnu - i\left[\Aenu - \Aznu \right]\cdot\Bmu \\
\label{231e}
\Gsmnu &= \nmu \Bsnu - \nnu \Bsmu - i\left[\Aemu - \Azmu \right]\cdot\Bsnu + i\left[\Aenu - \Aznu \right]\cdot\Bsmu \;.
\end{align}
\end{subequations}

The total field strength $\Fmnu$ \rf{213} appears now as the sum of the
``electromagnetic'' field strengths $\Famnu$ \rf{231b}\bs\rf{231c}, i.e.
\begin{equation}
\label{232}
\Fmnu = \Femnu + \Fzmnu \;,
\end{equation}
\noindent and therefore $\Fmnu$ is generated by the ``total potential'' $\Amu$
\begin{equation}
\label{233}
\Amu = \Aemu + \Azmu \;
\end{equation} 
\noindent in a strictly abelian way
\begin{equation}
\label{234}
\Fmnu = \nmu \Anu - \nnu \Amu \;.
\end{equation}
\noindent Recalling also the abelian character of the ``total'' Maxwell equations \rf{214} we see that the total objects of the system $\Amu,\Fmnu,\jmu$ obey a strictly abelian structure due to the ``total'' group $U(1)$, just as it is the case in conventional \kl quantum\kr\;electrodynamics.\\
\indent In contrast to these ``total'' properties, describing the $N$\bs particle system as a whole, the ``internal'' gauge structure is truly non\bs abelian on account of the exchange interactions. More concretely, when the current operator $\SMJmu$ \rf{224a} is decomposed as
\begin{equation}
\label{235}
\SMJmu = i\left\{ -\kamu \tau_a + \hsmu \chi - \hmu \bar{\chi} \right\}\;,
\end{equation}
\noindent the ``internal'' Maxwell equations \rf{225b} read in component form
\begin{subequations}
\begin{align}
\label{236a}
\nmo\Femnu + i\left[\Bmo\Gsmnu - \Bsmo\Gmn\right] &= -4\pi\as\kenu \\
\label{236b}
\nmo\Fzmnu - i\left[\Bmo\Gsmnu - \Bsmo\Gmn\right] &= -4\pi\as\kznu \\
\label{236c}
\nmo\Gmn + i\left[\Aemo - \Azmo\right]\Gmn - i\Bmo\left[\Femnu - \Fzmnu\right] &= 4\pi\as\hsnu \\
\label{236d}
\nmo\Gsmnu - i\left[\Aemo - \Azmo\right]\Gsmnu + i\Bsmo\left[\Femnu - \Fzmnu\right] &= 4\pi\as\hnu \;.
\end{align}
\end{subequations}
\noindent Adding here both equations \rf{236a} and \rf{236b} yields just the ``total'' Maxwell equations \rf{214} with the total current $\jmu$ now being found in terms of the electromagnetic single\bs particle currents $\kamu$ as
\begin{equation}
\label{237}
\jmu = \kemu + \kzmu \;.
\end{equation}
\indent Finally, one may also consider the non\bs abelian ``charge conservation''. Observe here, that the curvature $\MFmnu$ must obey the following bundle identity \kl valid in any space\bs time with vanishing torsion\kr
\begin{equation}
\label{238}
\MDmo\MDno\MFmnu \equiv 0 \;.
\end{equation}
\noindent Therefore differentiate once more the internal Maxwell equations \rf{225b} and find the following source equation in operator form
\begin{equation}
\label{239}
\MDmo \SMJmu \equiv 0 \;.
\end{equation}
\noindent Written in component form, this source relation reads
\begin{subequations}
\begin{align}
\label{240a}
\nmo\kemu &= i\left[\Bmo\hmu - \Bsmo\hsmu\right] \\
\label{240b}
\nmo\kzmu &= -i\left[\Bmo\hmu - \Bsmo\hsmu\right] \\
\label{240c}
\nmo\hmu &= i\left[\Aemo - \Azmo\right]\hmu + i\Bsmo\left[\kemu - \kzmu\right] \\
\label{240d}
\nmo\hsmu &= -i\left[\Aemo - \Azmo\right]\hsmu - i\Bmo\left[\kemu - \kzmu\right] \;.
\end{align}
\end{subequations}
\noindent As a consistency test, one may add up the first two source relations \rf{240a}\bs\rf{240b} in order to find just the former charge conservation law \rf{27} for the total current $\jmu$ \rf{237}.\\
\indent Clearly, the crucial point with this non\bs abelian gauge structure of RST is now whether one can identify certain experimental facts which support the non\bs abelian hypothesis. Since the latter aims at the \emph{exchange} interactions of many\bs particle systems, one may consider the stationary states of bound $N$\bs particle systems \kl e.g. many\bs electron atoms\kr\;which occur in form of entangled states implying the emergence of ``exchange energy''. Or in other words, verification of the RST predictions may be obtained by inspection of the atomic energy level systems, which are determined to a considerable extent by the exchange interactions. In order to prepare such a test of RST, it is very instructive to consider first an appropriate energy functional \kl$\ET$, say\kr\; for the RST field configurations.

\subsection{Energy Functional}

Within the context of a relativistic field theory, one would like to define the energy $\ET$ as the space integral over the time component $\TToo$ of an appropriate energy\bs momentum tensor $\TTmnu$. Equation \rf{216} suggests to take for $\TTmnu$ the sum of the internal part $\STmnu$ of the system and the interaction part $\esTmnu$ with respect to the external source
\begin{equation}
\label{241}
\TTmnu = \STmnu + \esTmnu \;.
\end{equation}
Since every $N$\bs particle system is composed of matter and gauge fields, mediating the interactions of matter, one will build up the internal density $\STmnu$ by a Dirac matter part $\DTmnu$ and a gauge part $\GTmnu$:
\begin{equation}
\label{242}
\STmnu = \DTmnu + \GTmnu \;.
\end{equation}
\noindent Thus the total density $\TTmnu$ \rf{241} of the $N$\bs particle system consists of three parts
\begin{equation}
\label{243}
\TTmnu = \DTmnu + \GTmnu + \esTmnu \;.
\end{equation}
\noindent Consequently, the total energy $\ET$
\begin{equation}
\label{244}
\ET = \vint\:\TToo\vvr 
\end{equation}
\noindent will also be built up by three constituents
\begin{equation}
\label{245}
\ET = \ED + \EG + \Ees \;,
\end{equation}
\noindent with the self\bs evident definitions
\begin{subequations}
\begin{align}
\label{246a}
\ED &= \vint\:\DToo\vvr \\
\label{246b}
\EG &= \vint\:\GToo\vvr \\
\label{246c}
\Ees &= \vint\:\esToo\vvr \;.
\end{align}
\end{subequations}
\indent Here the simplest part is $\Ees$ \rf{246c}, because the corresponding energy\bs momentum density $\esTmnu$ is in a bilinear way composed of the the external field strength $\exFmnu$ \rf{29} and the total field strength $\Fmnu$ \rf{213}:
\begin{equation}
\label{247}
\esTmnu = - \frac{\hbar c}{4\pi\as} \left\{ {}^\mathrm{(ex)}F_{\mu\lambda}\cdot\Fnulo + F_{\mu\lambda} \cdot \exFnulo - \frac{1}{2}g_{\mu \nu}\exFsulu\cdot\Fsolo \right\} \;.
\end{equation}
\noindent Consequently, the external interaction energy $\Ees$ \rf{246c} is given by
\begin{equation}
\label{248}
\Ees = \frac{\hbar c}{4\pi\as} \vint\: \left\{\vec{E}_{\mathrm{ex}} \cdot \E + \vec{H}_{\mathrm{ex}} \cdot \Hv \right\} \doteqdot \Eees + \Emes \;,
\end{equation}
\noindent where the three\bs vectors $\E\vvr = \left\{ E^j \right\}$ and $\Hv\vvr = \left\{ H^j \right\}$ of electric and magnetic field strengths are given in terms of the components of the field strength tensor $\Fmnu$ \rf{232} by
\begin{subequations}
\begin{align}
\label{249a}
E^j &\doteqdot F_{0 j} \\
\label{249b}
H^j&\doteqdot \frac{1}{2} \varepsilon^{j k}_{\;\;\;\;l} F_k^{\;\;l} \;,\text{etc.}
\end{align}
\end{subequations}
\noindent If the external fields $\vec{E}_{\mathrm{ex}}$, $\vec{H}_{\mathrm{ex}}$ are adopted to be constant \kl as it is usually done for treating the Zeeman and Stark effects\kr, the external energy $E_{es}$ \rf{248} becomes infinite and one has to substract the infinite contribution due to the external source generating the homogeneous fields $\vec{E}_{\mathrm{ex}},\vec{H}_{\mathrm{ex}}$ \kl see the discussion below equations \rf{334} and \rf{531}\kr. \\
\indent Next, considering the gauge field density $\GTmnu$, one finds this object to appear as the difference of the electromagnetic energy\bs momentum density $\RTmnu$ and the exchange density $\CTmnu$
\begin{equation}
\label{250}
\GTmnu = \RTmnu - \CTmnu 
\end{equation}
\noindent with the energy\bs momentum content of the electromagnetic modes being given by
\begin{equation}
\label{251}
\RTmnu = -\frac{\hbar c}{4\pi\as}\left\{\Femlu\Fznulo + \Fzmlu\Fenulo -\frac{1}{2}g_{\mu \nu} \Fesulu\Fzsolo\right\} \;,
\end{equation}
\noindent and similarly for the exchange modes of the gauge field system
\begin{equation}
\label{252}
\CTmnu = -\frac{\hbar c}{4\pi\as}\left\{\Gml\Gsnulo + \Gsmlu\Gnulo -\frac{1}{2}g_{\mu \nu}\Gsslu\Glso\right\} \;.
\end{equation}
\noindent Naturally, according to this subdivision, the gauge field energy $\EG$ \rf{246b} is also split up into two parts
\begin{equation}
\label{253}
\EG = \ER - \EC \;,
\end{equation}
\noindent where the electromagnetic energy $\ER$ is given by
\begin{equation}
\label{254}
\ER = \vint\:\RToo\vvr = \frac{\hbar c}{4\pi\as} \vint\:\left\{\E_1\cdot\E_2 + \Hv_1\cdot\Hv_2\right\} \;,
\end{equation}
\noindent and similarly for the exchange energy $\EC$
\begin{equation}
\label{255}
\EC = \vint\:\CToo\vvr = \frac{\hbar c}{4\pi\as}\vint\:\left\{\vec{X}^{*}\cdot\vec{X} + \vec{Y}^{*}\cdot\vec{Y}\right\} \;.
\end{equation}
\noindent Clearly, the ``electric'' and ``magnetic'' exchange three\bs vectors $\vec{X}=\left\{X^j\right\}$ and $\vec{Y}=\left\{Y^j\right\}$ are defined analogously to their electromagnetic counterparts \rf{249a}\bs\rf{249b}
\begin{subequations}
\begin{align}
\label{256a}
X^j &\doteqdot G_{0j} \\
\label{256b}
Y^j &\doteqdot \frac{1}{2} \varepsilon^{j k l}G_{kl} \;.
\end{align}
\end{subequations}
\indent Now it is evident that both the electromagnetic energy $\ER$ \rf{254} and the exchange energy $\EC$ \rf{255} may be split up further into their ``electric'' and ``magnetic'' constituents, i.e. we put
\begin{subequations} 
\begin{align}
\label{257a}
\ER &= \ERe + \ERm \\
\label{257b}
\EC &= \ECh + \ECg \;,
\end{align}
\end{subequations}
\noindent with the ``electric'' parts being defined in an obvious way through
\begin{subequations}
\begin{align}
\label{258a}
\ERe &\doteqdot \frac{\hbar c}{4\pi\as}\vint\:\E_1\cdot\E_2 \\
\label{258b}
\ECh &\doteqdot \frac{\hbar c}{4\pi\as}\vint\:\vec{X}^{*}\cdot\vec{X} \;,
\end{align}
\end{subequations}
\noindent and similarly for the ``magnetic'' parts
\begin{subequations}
\begin{align}
\label{259a}
\ERm &\doteqdot \frac{\hbar c}{4\pi\as}\vint\:\Hv_1\cdot\Hv_2 \\
\label{259b}
\ECg &\doteqdot \frac{\hbar c}{4\pi\as}\vint\:\vec{Y}^{*}\cdot\vec{Y} \;.
\end{align}
\end{subequations}
\noindent The important point with this subdivision into ``electric'' and ``magnetic'' parts of the gauge energy is, that the ``electric'' contributions $\ERe$ and $\ECh$ will turn out to be much larger than their ``magnetic'' counterparts $\ERm$ and $\ECg$, so that the latter ones may be treated as small perturbations. The main effect is therefore due to the ``electric'' contributions and represents thus by itself an acceptable first approximation to the experimental data \kl{\it Electrostatic Approximation}\kr. In this sense we may rearrange the gauge field energy $\EG$ \rf{253} as
\begin{equation}
\label{260}
\EG = \ERe - \ECh + \left(\ERm - \ECg\right) \;,
\end{equation}
\noindent so that the ``magnetic'' contributions \kl in brackets\kr\;can be omitted for the electrostatic approximation. This approximation is then based solely on the truncated gauge field energy $\EGeh$
\begin{equation}
\label{261}
\EGeh \doteqdot \ERe - \ECh
\end{equation}
\noindent and thus misses the magnetic energy $\EGmg$
\begin{equation}
\label{262}
\EGmg \doteqdot \ERm - \ECg \;,
\end{equation}
\noindent which will be considered as the ``magnetic correction'' of the electrostatic energy $\EGeh$ \rf{261}.
\indent Finally, the matter energy $\ED$ \rf{246a} has to be considered. The corresponding energy\bs momentum density $\DTmnu$ carried by the wave function $\Psi$ is given by
\begin{equation}
\label{263}
\DTmnu = \frac{i\hbar c}{4} \left\{\bar{\Psi}\Gamma_{\mu}\left(\MDnu\Psi\right) - \left(\MDnu\bar{\Psi}\right)\Gamma_{\mu}\Psi + \bar{\Psi}\Gamma_{\nu}\left(\MDmu\Psi\right) - \left(\MDmu\bar{\Psi}\right)\Gamma_{\nu}\Psi\right\} \;.
\end{equation}
\noindent Intuitively, one expects the associated matter energy $\ED$ to consist of the rest mass energy $2Mc^2$ of both particles \kl take for the moment $N=2$\kr\;plus their kinetic energy $E_{kin}$, i.e. one expects
\begin{equation}
\label{264}
\ED = 2Mc^2 + \sum\limits_{a=1}^{2} E_{kin\:(a)} \;.
\end{equation}  
Indeed, this result can actually be deduced in the non\bs relativistic limit \cite{Hu}. In
order to have a brief sketch of the proof, eliminate the time derivative $\MDou\Psi$ from
the matter energy density $\DToo$
\begin{equation}
\label{265}
\DToo = \frac{i \hbar c}{2} \left\{\bar{\Psi}\Gamma_0\left(\MDou\Psi\right) - \left(\MDou\bar{\Psi}\right)\Gamma_0\Psi\right\}
\end{equation}
by means of the Dirac equation \rf{222} and find
\begin{equation}
\label{266}
\DToo = Mc^2\bar{\Psi}\Psi + \frac{i\hbar c}{2}\left\{\left(\MDju\bar{\Psi}\right)\Gamma^j\Psi - \bar{\Psi}\Gamma^j\MDju\Psi\right\} \;.
\end{equation}
Integrating now over all three\bs space and carrying out the non\bs relativistic limit in
a consistent way yields just the claimed result \rf{264}, where the correspondence of rest
mass terms and kinetic energy terms should be obvious by comparing both equations \rf{264}
and \rf{266}.

Observe also, that for the stationary bound field configurations \kl as solutions
of the mass eigenvalue equations\kr\;the mass eigenvalues should appear somewhere in the
energy functional. For one\bs particle systems, one may even expect that the field energy
$\ET$ has to coincide with the mass\bs energy eigenvalue \kl$M_{*}c^2$,say\kr\;of the
bound particle \cite{Sch}:

\begin{equation}
\label{267}
\ET = M_{*}c^2 \;.
\end{equation}
The relationship between the field energy $\ET$ and the mass eigenvalues for few\bs
particle systems has now to be studied in more detail, together with their dependence on
the ``electric'' and ``magnetic'' interactions.


\section{Electrostatic Approximation}

The relative magnitude of the ``electric'' and ``magnetic'' interparticle interactions in
a stationary bound system strongly depends on the value of the electromagnetic coupling
constant $\as$. In view of its smallness \kl$\as\approx\frac{1}{137}$\kr, the ``electric''
interactions clearly dominate the ``magnetic'' interactions, because their relative
magnitude is of the order $\left(z_{ex}\as\right)^2$, where $z_{ex}$ is the number of \kl
positive\kr\;charge units located at the center of the binding force, e.g. the nucleus
\cite{Ve2}. Therefore, at least for the light atoms \kl$z_{ex} \le 10,...,20$, say\kr, one
may speculate that the complete neglection of the interactions of the ``magnetic'' type
\kl''electrostatic approximation''\kr\;yields a first useful estimate of the relativistic
energy levels of the bound system. The reason for this expectation is that, as a
consequence of their dominance over the magnetic interactions, the electric interactions
will also contribute the main part of the relativistic effects. If this presumption is
true \kl to be verified readily\kr, one will not only be able to treat the magnetic
interactions as a small perturbation but furthermore the non\bs relativistic limit of this
perturbation will be all that is needed for taking into account the magnetic effect to
first order. In chapter VI we will demonstrate that this apparently rough approximation
scheme gets the RST predictions closer to the experimental data of ref~\cite{MES} ending
up with less than $0,5\%$ of relative deviation \kl see table~II\kr.
\\
\\
\indent
Thus we will proceed as follows:
\begin{itemize}
\item[(i)] elaborating in this chapter the relativistic stationary field configurations as far as they are needed for the electrostatic approximation
\end{itemize}
\noindent and     
\begin{itemize}
\item[(ii)] computing the magnetic corrections in the {\it non\bs relativistic} limit in the subsequent  chapters.
\end{itemize}

\subsection{Stationary States}

The stationary bound states are not completely time\bs independent but have such a temporal behaviour that the observable quantities of the theory \kl i.e. eigenvalues and densities of charge, current, energy etc.\kr\;become truly time\bs independent. Thus the wave functions of the considered two\bs particle system \kl$N=2$\kr\; vary with time as
\begin{subequations}
\begin{align}
\label{31a}
\psi_1\vvrt &= \exp\left[-i\frac{M_1 c^2}{\hbar}t\right]\cdot\psi_1\vvr \\
\label{31b}
\psi_2\vvrt &= \exp\left[-i\frac{M_2 c^2}{\hbar}t\right]\cdot\psi_2\vvr \;,
\end{align}
\end{subequations}
\noindent where $M_a$  \kl$a=1,2$\kr\;denote the {\it mass eigenvalues}. This stationary form of the wave functions implies the time\bs independence of the electromagnetic objects, i.e. the currents $k^a_{\;\;\mu}$, potentials $A^a_{\;\;\mu}$ and field strengths $\vec{E}_a, \vec{H}_a$:
\begin{subequations}
\begin{align}
\label{32a}
\left\{k^a_{\;\;\mu}\right\} &\Rightarrow \left\{\ako;\kva\right\} \\
\label{32b}
\left\{A^a_{\;\;\mu}\right\} &\Rightarrow \left\{ \aAo; \Ava \right\} \\
\label{32c}
\vec{E}_a &\Rightarrow \Eva \\
\label{32d}
\vec{H}_a &\Rightarrow \Hva \;,
\end{align}
\end{subequations}
\noindent whereas their exchange counterparts adopt the following time dependence:
\begin{subequations}
\begin{align}
\label{33a}
h_{\mu}\vvrt &= \exp\left[i\frac{M_1-M_2}{\hbar} c^2 t\right] \cdot \hm \\
\label{33b}
h_{\mu}^{*}\vvrt &= \exp\left[-i\frac{M_1-M_2}{\hbar} c^2 t\right] \cdot \hms \\
\label{33c}
B_{\mu}\vvrt &= \exp\left[-i\frac{M_1-M_2}{\hbar} c^2 t\right] \cdot \Bmu\vvr \\
\label{33d}
B_{\mu}^{*}\vvrt &= \exp\left[i\frac{M_1-M_2}{\hbar} c^2 t\right] \cdot \Bsmu\vvr \\
\label{33e}
\vec{X}\vvrt &= \exp\left[-i\frac{M_1-M_2}{\hbar} c^2 t\right] \cdot \X \\
\label{33f}
\vec{Y}\vvrt &= \exp\left[-i\frac{M_1-M_2}{\hbar} c^2 t\right] \cdot \Y \;.
\end{align}
\end{subequations}
\indent The relativistic eigenvalue problem for the determination of the mass eigenvalues $M_a$ is obtained now by substituting the stationary form of the wave functions into the two\bs particle Dirac equation \rf{222} which yields the following coupled system for the spatial parts $\psi_a\vvr$ of the stationary wave functions $\psi_a\vvrt$ \rf{31a}\bs\rf{31b}:
\begin{subequations}
\begin{align}
\label{34a}
i\gamma^j\MBDju \psi_1\vvr + \left[ \exAou + \zAou \right] \gamma_0 \psi_1\vvr + \Bou \gamma_0\psi_2\vvr &= \frac{M-M_1\gamma_0}{\hbar}\,c\,\psi_1\vvr \\
\label{34b}
i\gamma^j\MBDju \psi_2\vvr + \left[ \exAou + \eAou \right] \gamma_0 \psi_2\vvr + \Bous \gamma_0\psi_1\vvr &= \frac{M-M_2\gamma_0}{\hbar}\,c\,\psi_2\vvr \;.
\end{align}
\end{subequations}
\noindent Here the time derivatives of the wave functions have brought in the mass eigenvalues $M_a$, and the remaining spatial derivations $\MBDju$ are defined as follows:
\begin{subequations}
\begin{align}
\label{35a}
\MBDju \psi_1\vvr &\doteqdot \pju \psi_1\vvr - i\left[\exAju + \zAju\right] \psi_1\vvr - i\Bju\psi_2\vvr \\
\label{35b}
\MBDju \psi_2\vvr &\doteqdot \pju \psi_2\vvr - i\left[\exAju + \eAju\right] \psi_2\vvr - i\Bjus\psi_1\vvr \;.
\end{align}
\end{subequations}
\indent Now, by its very definition, the {\it electrostatic approximation} consists in neglecting the interactions of the ``magnetic'' type, i.e. one puts to zero all three\bs vector potentials $\Ava, \Bv$ and thereby ends up with the following truncated system:
\begin{subequations}
\begin{align}
\label{36a}
i\gamma^j\pju\psi_1\vvr + \left[\exAou + \zAou\right]\gamma_0\psi_1\vvr + \Bou \gamma_0 \psi_2\vvr &= \frac{M-\tilde{M}_1\gamma_0}{\hbar}\,c\,\psi_1\vvr \\
\label{36b}
i\gamma^j\pju\psi_2\vvr + \left[\exAou + \eAou\right]\gamma_0\psi_2\vvr + \Bous \gamma_0 \psi_1\vvr &= \frac{M-\tilde{M}_2\gamma_0}{\hbar}\,c\,\psi_2\vvr \;.
\end{align}
\end{subequations}
\noindent Logically, both Dirac spinors $\psi_a\vvr$ \kl$a=1,2$\kr\;do couple here exclusively to the remaining time components $\aAo,\Bou$ of the corresponding four\bs vector potentials $A^a_{\;\;\mu},B_{\mu}$. Therefore we have to complement the RST\bs Dirac system \rf{36a}\bs\rf{36b} merely by the Poisson equations for these time components, which of course have to be deduced from the Maxwell equations \rf{236a}\bs\rf{236d} by neglecting again the spatial components of the four\bs vector potentials:
\begin{subequations}
\begin{align}
\label{37a}
\Delta \aAo &= 4\pi\as \ako \\
\label{37b}
\Delta \Bou &= -4\pi\as \hos \;.
\end{align}
\end{subequations}
\noindent Observing here that the currents $k^a_{\;\;\mu},h_{\mu}$ are generated by the wave functions through
\begin{subequations}
\begin{align}
\label{38a}
k^a_{\;\;\mu}\vvr &= \bar{\psi}_a\vvr\gamma_{\mu}\psi_a\vvr \\
\label{38b}
h_{\mu}\vvr &= \bar{\psi}_1\vvr\gamma_{\mu}\psi_2\vvr \;,
\end{align}
\end{subequations}
\noindent one arrives at a coupled but closed Dirac\bs Poisson system \rf{36a}\bs\rf{37b} whose solutions $\left\{\tilde{\psi}_a\vvr,\tilde{M}_a\right\}$ constitute what we consider to be the ``electrostatic approximation''. \\
\indent Subsequently, we want to carry out a test of the usefulness of RST by opposing its theoretical predictions for the two\bs particle ground\bs state in the Coulomb potential
\begin{equation}
\label{39}   
\exAou = z_{ex}\frac{\as}{|\vec{r}|}
\end{equation}
\noindent to the experimental data. Since the ground\bs state is a member of the para\bs sytem with the highest possible symmetry, we try the following ansatz for the Dirac spinors $\tilde{\psi}_a\vvr$:
\begin{equation}
\label{310}
\tilde{\psi}_a\vvr = \begin{pmatrix} \afp\vvr \\ \afm\vvr \end{pmatrix}
\end{equation}
\noindent with the first particle having spin up, i.e. the two component Pauli spinors are chosen as
\begin{subequations}
\begin{align}
\label{311a}
\efp\vvr &= \eRp \cdot \zeoe \\
\label{311b}
\efm\vvr &= -i\, \eRm \cdot \zeee \;.
\end{align}
\end{subequations}
\noindent Here the one\bs particle eigenspinors $\zjlm$ of the angular momentum operators obey the relations \cite{Me,Th}
\begin{subequations}
\begin{align}
\label{312a}
\vec{J}^2 \,\zjlm &= j(j+1)\hbar^2\cdot \zjlm \\
\label{312b}
J_z \,\zjlm &= m\hbar \cdot \zjlm \\
\label{313c}
\vec{L}^2 \,\zjlm &= l(l+1)\hbar^2 \cdot \zjlm \\
\label{313d}
\vec{S}^2 \,\zjlm &= \frac{1}{2} \left(\frac{1}{2} + 1\right) \hbar^2 \cdot \zjlm \;,
\end{align}
\end{subequations}
\noindent with the composition law for spin\bs$\frac{1}{2}$ particles
\begin{equation}
\label{313}
j = l\pm\frac{1}{2} \;.
\end{equation}
\noindent Since the two spins are anti\bs parallel in the ground\bs state, we put for the second particle
\begin{subequations}
\begin{align}
\label{314a}
\zfp\vvr &= \zRp \cdot \zeome \\
\label{314b}
\zfm\vvr &= -i\,\zRm \cdot \zeeme \;.
\end{align}
\end{subequations}
\indent Now, for the ground state, it is reasonable to assume that the radial parts $\aRpm$ of both wave functions $\tilde{\psi}_a\vvr$ are identical
\begin{subequations}
\begin{align}
\label{315a}
\eRp &\equiv \zRp \doteqdot \Rp \\
\label{315b}
\eRm &\equiv \zRm \doteqdot \Rm \;,
\end{align}
\end{subequations}
\noindent and similarly for the mass eigenvalues
\begin{equation}
\label{316}
\tilde{M}_1 = \tilde{M}_2 \doteqdot \tMo \;.
\end{equation}
\noindent Thus the original RST eigenvalue system \rf{36a}\bs\rf{36b} becomes simplified to two coupled equations for the two radial ansatz functions $\Rpm$:
\begin{subequations}
\begin{align}
\label{317a}
\frac{{\rm d}\Rp}{{\rm d}r} + \left[\exAour + A_0(r)\right]\cdot \Rm &= - \frac{\tMo + M}{\hbar}\,c\cdot \Rm \\
\label{317b}
\frac{{\rm d}\Rm}{{\rm d}r} +\frac{2}{r}\Rm - \left[\exAour + A_0(r)\right]\cdot \Rp &=  \frac{\tMo - M}{\hbar}\,c\cdot \Rp \;.
\end{align}
\end{subequations}
\indent Observe here that for the para\bs ansatz \rf{314a}\bs\rf{314b} the time component $\ho$ of the exchange current $\hmu$ \rf{38b} vanishes \kl$\ho\equiv 0$\kr\;so that one can take a vanishing exchange potential \kl$\Bou\equiv 0$\kr\;as solution of the Poisson equation \rf{37b}. Therefore the radial functions $\Rpm$ do couple only to the electrostatic potentials ${}^{(a)}\!A_0(r)$ which of course now have to be identical, too
\begin{equation}
\label{318}
{}^{(1)}\!A_0(r) \equiv {}^{(2)}\!A_0(r) \doteqdot A_0(r)
\end{equation}
\noindent and obey the Poisson equation \rf{37a}
\begin{equation}
\label{319}
\Delta A_0(r) = 4\pi\as k_0(r) \;,
\end{equation}
\noindent with the coinciding electric charge densities $\ako$ \rf{32a} being given by
\begin{equation}
\label{320}
{}^{(1)}\!k_0(r) \equiv {}^{(2)}\!k_0(r) \doteqdot k_0(r) = \frac{\left(\Rp\right)^2 + \left(\Rm\right)^2}{4\pi} \;.
\end{equation}
\noindent Thus the formal solution of the Poisson equation \rf{319} is found to be
\begin{equation}
\label{321}
A_0(r) = -\frac{\as}{4\pi} \vints \: \frac{\left(R_{+}(r')\right)^2 + \left(R_{-}(r')\right)^2}{\brrs} \;.
\end{equation}

\subsection{Non\bs relativistic Limit}

Since the electrostatic approximation is required to take fully into account the relativistic effects, the coupled system \rf{317a}\bs\rf{320} has to be solved numerically and then yields the basis for the RST results $\Delta E^{(e)}_{RST}$ \kl third row of table~I\kr. However, for the calculation of the magnetic corrections we may restrict ourselves to the lowest\bs order approximation of the solutions $\tilde{\psi}_a\vvr$. This means that we
\begin{itemize}
\item[(i)] neglect the interaction between both particles \kl i.e. putting $A_0(r) \Rightarrow 0$\kr
\end{itemize}
\noindent and
\begin{itemize}
\item[(ii)] we are satisfied with the non\bs relativistic approximation of the corresponding solutions $\Rpm$.
\end{itemize}

\indent Now, by virtue of the first assumption, the interactive system \rf{317a}\bs\rf{317b} becomes decoupled and truncated to the simpler form
\begin{subequations}
\begin{align}
\label{322a}
\frac{{\rm d}\Rp}{{\rm d}r} + \exAour \cdot \Rm &= - \frac{\Ms + M}{\hbar}\,c\cdot \Rm \\
\label{322b}
\frac{{\rm d}\Rm}{{\rm d}r} +\frac{2}{r}\Rm - \exAour \cdot \Rp &=  \frac{\Ms - M}{\hbar}\,c\cdot \Rp \;,
\end{align}
\end{subequations}
\noindent whose solutions are given by \cite{Gr,Gr2}
\begin{subequations}
\begin{align}
\label{323a}
\Rp &= N_{*}\sqrt{M+\Ms}\,r^{\nu} \exp\left[-\frac{z_{ex}r}{\aB}\right] \\
\label{323b}
\Rm &= N_{*}\sqrt{M-\Ms}\,r^{\nu} \exp\left[-\frac{z_{ex}r}{\aB}\right] \\
\nonumber
\Big( \aB  &= \frac{\hbar^2}{Me^2} \ldots\text{Bohr radius}\Big).
\end{align}
\end{subequations}
\noindent Here the parameter $\nu$ is given by
\begin{equation}
\label{324}
\nu = -1 + \sqrt{1 - \left(z_{ex}\as\right)^2} \;.
\end{equation}
\noindent Furthermore the mass eigenvalue $\tilde{M}_0$ of the coupled two\bs particle system \rf{317a}\bs\rf{317b} degenerates to the well\bs known one\bs particle result $\Ms$ \cite{Gr,Gr2}
\begin{equation}
\label{325}
\Ms = M\sqrt{1 - \left(z_{ex}\as\right)^2} \;;
\end{equation}
\noindent and finally the normalization constant $N_{*}$ is computed via the relativistic normalization condition
\begin{equation}
\label{326}
\vint\:k_0\vvr = 1
\end{equation}
\noindent as
\begin{equation}
\label{327}
N_{*}^2 = \frac{1}{2M}\frac{\left(\frac{2 z_{ex}}{\aB}\right)^{3+2\nu}}{\Gamma(3+2\nu)} \;.
\end{equation}
\indent The second assumption \kl of non\bs relativistic approximation\kr\;says that the radial function $\Rm$ \rf{323b} is neglected against $\Rp$ and for the latter function \rf{323a} one takes its non\bs relativistic limit $\overset{\circ}{R}_{+}(r)$, i.e.
\begin{equation}
\label{328}
\Rp \Rightarrow \overset{\circ}{R}_{+}(r) = 2\sqrt{\left(\frac{z_{ex}}{\aB}\right)^3}\exp\left[-\frac{z_{ex}r}{\aB}\right]
\end{equation}
\noindent which obeys the non\bs relativistic version of the normalization condition \rf{326}
\begin{equation}
\label{329}
\int\limits_{0}^{\infty}{\rm d}r\:r^2\left(\overset{\circ}{R}_{+}(r)\right)^2 = 1 \;.
\end{equation}
\noindent Clearly, this non\bs relativistic limit $\overset{\circ}{R}_{+}(r)$ is the ground\bs state solution of the ordinary Schr\"odinger equation
\begin{gather}
\label{330}
-\frac{\hbar^2}{2M}\Delta\psi\vvr - \hbar c\exAou\psi\vvr = E_0\psi\vvr \\
\nonumber
\Big( E_0 = -\frac{z_{ex}^2e^2}{2\aB}\Big)
\end{gather}
\noindent and therefore it may appear somewhat amazing that the ordinary Schr\"odinger equation is sufficient in order to calculate the magnetic corrections in the lowest\bs order approximation. However, as we shall readily see, relativity nevertheless enters the ``magnetic'' exchange energy, namely in form of the exchange current $\hv$ which is built up by the radial functions $\aRpm$ through the combination
\begin{equation}
\label{331}
\MRp(r) \doteqdot \eRp\cdot\zRm + \zRp\cdot\eRm \;,
\end{equation}
\noindent see equation \rf{558} below. Observing here the coincidence of the radial functions $\aRpm$ for the two\bs particle ground\bs state \rf{315a}\bs\rf{315b}, one deduces from the non\bs relativistic approximations \rf{323a}\bs\rf{323b} the corresponding approximation for $\MRp(r)$ \rf{331} as
\begin{equation}
\label{332}
\MRp(r) \Rightarrow 4\left(z_{ex}\as\right)\frac{z_{ex}^3}{\aB^3}\exp\left[-2\frac{z_{ex}r}{\aB}\right] \;.
\end{equation}

\subsection{Mass Eigenvalues $M_a$ and Field Energy $\ET$}

Intuitively, one should suppose that the total energy $\ET$ \rf{245}is related to the mass eigenvalues $M_a$ in some way. Indeed, this desired relation can easily be obtained by inserting the stationary wave functions $\psi_a\vvr$ \kl see the right\bs hand sides of \rf{31a}\bs\rf{31b}\kr, potentials $A^a_{\;\;\mu}$ \rf{32b} and $B_{\mu}$ \rf{33c}\bs\rf{33d} into the matter density $\DToo$ \rf{265} and integrating over all three\bs space. In this way, one finds quite generally the following result for the matter energy $\ED$ \rf{246a} in terms of the mass eigenvalues $M_a$
\begin{eqnarray}
\label{333}
\ED &=& \hat{z}_1\cdot M_1c^2 + \hat{z}_2\cdot M_2c^2 - \Eees \\
\nonumber
 & & {}+\hbar c \vint\:\left\{\eAo\cdot\zko + \zAo\cdot\eko\right\} \\
\nonumber
 & & {}+\hbar c \vint\:\left\{\Bou\cdot\ho + \Bous\cdot\hos\right\} \;.
\end{eqnarray}
\noindent Here it is presumed that the external interaction energy of the electric type $\Eees$ \rf{248} can be recasted into the following form by means of Gau\ss' integral theorem and the Poisson equations \rf{37a}:
\begin{equation}
\label{334}
\Eees \Rightarrow -\hbar c \vint\:\exAou\left\{\eko + \zko\right\} \;.
\end{equation}
\noindent This presumption implies that the external field $\vec{E}_{ex}\vvr$ is well\bs localized; if this is not true \kl e.g. by considering a homogeneous field $\vec{E}_{ex}$\kr, the Gau\ss' surface term becomes infinite at infinity and has to be omitted. The reason is that such an infinite surface term has to be considered as part of the \kl infinite\kr\;energy of the external non\bs localized source generating the homogeneous field $\vec{E}_{ex}$. Moreover, the former normalization condition \rf{326} upon the wave functions $\psi_a\vvr$, being adequate for the electrostatic approximation, has to be generalized in the presence of magnetic interactions to
\begin{equation}
\label{335}
\vint\:\ako = \hat{z}_a \;,
\end{equation}
\noindent with the real numbers $\hat{z}_a$ still close to unity \kl for the relativistic normalization of wave functions in the general case see ref.\cite{Ve2,Hu}\kr. \\
\indent Now one substitutes the present result for the matter energy $\ED$ \rf{333} into the total field energy $\ET$ \rf{245} which then reappears in the following form:
\begin{eqnarray}
\label{336}
\ET &=& \hat{z}_1\cdot M_1c^2 + \hat{z}_2\cdot M_2c^2 + \Emes \\
\nonumber 
 & & {}+ \ERe + \hbar c\vint\:\left\{\eAo\cdot\zko + \zAo\cdot\eko\right\} \\
\nonumber
 & & {}- \ECh + \hbar c\vint\:\left\{\Bou\cdot\ho + \Bous\cdot\hos\right\} \\
\nonumber
 & & {}+ \ERm - \ECg \;.
\end{eqnarray}
\noindent However, the gauge objects emerging here \kl namely the potentials $\aAo$ and the charge densities $\ako$\kr\;are not independent, but rather are connected by the Poisson equations \rf{37a}; and this establishes the following relationship among the electrostatic energy contributions
\begin{equation}
\label{337}
\ERe = -\frac{\hbar c}{2}\vint\:\left\{\eAo\cdot\zko + \zAo\cdot\eko\right\} \;.
\end{equation}
\noindent This result again comes about by applying Gau\ss' integral theorem to the internal energy functional $\ERe$ \rf{258a} and using also the Poisson equations \rf{37a}. Clearly, there exists an analogous relationship for the ``electric'' exchange energy $\ECh$ \rf{258b}, namely
\begin{equation}
\label{338}
\ECh = \frac{\hbar c}{2}\vint\:\left\{\Bou\cdot\ho + \Bous\cdot\hos\right\} + \frac{\hbar c}{4\pi\as\aM^2}\vint\:\Bv\cdot\Bvs \;,
\end{equation}
\noindent which will be discussed in greater detail below. By means of these relationships the total energy $\ET$ \rf{336} can now be simplified to the plausible result
\begin{equation}
\label{339}
\ET = \hat{z}_1\cdot M_1c^2 + \hat{z}_2\cdot M_2c^2 + \Emes - \Delta\EGeh + \EGmg \;.
\end{equation}
\noindent Here the ``magnetic'' contribution $\EGmg$ \rf{262} reads in terms of the vector potentials $\Ava$ and $\Bv$
\begin{eqnarray}
\label{340}
\EGmg &=& -\frac{1}{2}\hbar c \vint\:\left\{\kve\cdot\Avz + \kvz\cdot\Ave + \hv\cdot\Bv + \hvs\Bvs\right\} \\
\nonumber
 & & {}- \frac{\hbar c}{4\pi\as\aM^2}\vint\:\Bvs\cdot\Bv \;,
\end{eqnarray}
\noindent and similarly its ``electric'' counterpart is defined through
\begin{eqnarray}
\label{341}
\Delta\EGeh &\doteqdot& \EGeh + \frac{\hbar c}{2\pi\as\aM^2}\vint\:\Bvs\cdot\Bv \\
\nonumber
 &=& - \frac{\hbar c}{2}\vint\:\left\{\eAo\cdot\zko + \zAo\cdot\eko + \Bou\cdot\ho + \Bous\hos\right\} \\
\nonumber
 & & {} +  \frac{\hbar c}{4\pi\as\aM^2}\vint\:\Bvs\cdot\Bv \;.
\end{eqnarray}
\indent This now is a very interesting result, because it provides us with much insight into the difference between the ``electric'' and the ``magnetic'' interactions. First observe that either of the two mass eigenvalues $M_a$ \kl$a=1,2$\kr\;already completely incorporates the ``electric'' plus ''magnetic'' interactions of one particle with the other, respectively.This can be most clearly seen from the original form of the eigenvalue equations \rf{34a}\bs\rf{34b}, namely by multiplying those equations by $\bar{\psi}_1$ or $\bar{\psi}_2$, resp., from the left and integrating over all three\bs space and finally resolving for the eigenvalues $M_a$. Therefore the sum of the mass\bs energy eigenvalues $\hat{z}_a\cdot M_a c^2$, occurring in the total energy $\ET$ \rf{339}, counts the electromagnetic interaction energy between both particles {\it twice} whereby the ``electric'' parts enter with a positive sign and the ``magnetic'' parts with a negative sign. Now, one of the pleasant features of the RST energy functional $\ET$ \rf{339} is that this double\bs counting is corrected by {\it subtraction} of the ``electric'' interaction part $\Delta\EGeh$ and {\it addition} of the ``magnetic'' interaction part $\EGmg$.The relatively good agreement of this theoretical picture with the experimental data \kl see table~I and~II\kr\;may be taken as a support of both the RST philosophy and of its practical consequences which will be considered readily. \\
\indent As far as the method of {\it electrostatic approximation} is concerned, it is self\bs suggesting to omit the magnetic terms in the total energy $\ET$ \rf{339}, quite similarly to the way the electrostatic eigenvalue system \rf{36a}\bs\rf{36b} was obtained from the general sytem \rf{34a}\bs\rf{34b} by omitting the ``magnetic'' potentials $\Ava$ and $\Bv$. This leads us to define the electrostatic approximation $\tilde{E}_T$ of the total $\ET$ by
\begin{equation}
\label{342}
\tilde{E}_T = \tilde{M}_1c^2 + \tilde{M}_2c^2 - \Delta\EGeh \;,
\end{equation}
\noindent where the mass eigenvalues $\tilde{M}_a$ are the eigenvalues of the truncated system \rf{36a}\bs\rf{36b} and the ``electric'' interaction functional $\Delta\EGeh$ is to be taken upon the solutions $\tilde{\psi}_a\vvr$ of this truncated system:
\begin{equation}
\label{342n}
\Delta\EGeh \Rightarrow -\frac{\hbar c}{2}\vint\:\left\{\eAou\cdot\zko + \zAou\cdot\eko + \Bou\cdot\ho + \Bos\cdot\hos\right\} \;.
\end{equation}
\noindent The corresponding numerical results for the ground\bs state interaction energy $\DERSTe$ are presented in the third column of table~I.

\subsection{Numerical Results}

Within the framework of the electrostatic approximation all prerequisites are now at hand in order to test the quality of this type of approximation. The numerical solution of the electrostatic eigenvalue system \rf{317a}\bs\rf{317b} provides us with the radial ground\bs state functions $\Rpm$ together with the mass eigenvalue $\tMo$. Consequently, these results may now be substituted into the electrostatic energy functional $\tilde{E}_T$ \rf{342}, whose value upon the calculated ground\bs state solutions $\left\{\Rpm;\tMo\right\}$ appears then as
\begin{equation}
\label{343n}
\tilde{E}_T \Rightarrow 2\tMo c^2 + \hbar c \vint\:A_0(r)\cdot k_0(r)  
\end{equation}
\noindent where the electrostatic potential $A_0(r)$ and the charge density $k_0(r)$ are given by equations \rf{320}\bs\rf{321}. Observe here also that the exchange potential $B_0(r)$ as a solution of the Poisson equation \rf{37b} vanishes because the exchange current $\hm$ \rf{38b} has a vanishing time\bs component \kl$\ho \equiv 0$\kr\;for the stationary field configurations being defined through equations \rf{311a}\bs\rf{311b} and \rf{314a}\bs\rf{314b}.\\
\indent Now the problem is how to judge of the present RST prediction $\tET$ \rf{343n} for
the two\bs electron ground\bs state energy when this is not directly observable. However, what can be {\it directly} measured is the interaction energy $\DEexp$ between the two electrons in the ground\bs state. This was done for six highly ionized elements, ranging from germanium \kl$z_{ex}=32$\kr\;up to bismuth \kl$z_{ex}=83$\kr, see ref.\cite{MES}. On the other hand, this interaction energy of the two ground\bs state electrons arises within the framework of RST as the difference $\DERSTe$ of the ground\bs state energy $\tET$ \rf{343n} and the double value \kl$2\Ms c^2$\kr\;of the single\bs particle energy eigenvalues \rf{325}, i.e.
\begin{eqnarray}
\label{344n}
\DERSTe &\doteqdot& \tET - 2\Ms c^2 \\
\nonumber
 &= & {}2\left(\tMo - \Ms\right)c^2 + \hbar c \vint\:A_0(r)\cdot k_0(r)\;.
\end{eqnarray}
\noindent The comparison of this RST prediction $\DERSTe$ with the experimental values
$\Delta E_{exp}$ displays some very instructive features of both the electric and the
magnetic interactions \kl see table~I\kr.
\begin{table}[hhh!]
\begin{tabular}{c||c|c|c|c|c}
Element \kl$z_{ex}$\kr & $\Delta\Eexp$ [eV] & $\Delta\ERSTe$ [eV] & $\Delta =
\frac{\Delta\Eexp - \DERSTe}{\Delta\Eexp}$ [\%] & $f_{*}^2 $ & $\varepsilon_{*}$ \\ \hline\hline
Ge \kl32\kr & 562,5$\pm$1,6 & 553,0 & 1,7 & 0,297 & 16,8 \\ \hline
Xe \kl54\kr & 1027,2$\pm$3,5 & 974,3 & 5,1 & 0,295 & 16,58 \\ \hline
Dy \kl66\kr & 1341,6$\pm$4,3 & 1232 & 8,2 & 0,294 & 16,36 \\ \hline
W \kl74\kr & 1568$\pm$15 & 1423 & 9,3 & 0,247 & 16,18 \\ \hline
Bi \kl83\kr & 1876$\pm$14 & 1661 & 11,5 & 0,223 & 15,92
\end{tabular}\medskip\medskip
\caption{\label{T1} Comparison of experimental values $\Delta\Eexp$ \kl second column\kr\;\cite{MES} with the RST predictions $\DERSTe$ \rf{344n} \kl third column\kr\;for the ground\bs state interaction energy of helium\bs like ions. The last two columns display the geometric factor $\fs^2$ \rf{357n} for the magnetic interactions and the reference energy $\es$ \rf{350n} for the electric interactions. Both limit values $f_0^2=0,4$ \rf{643} and $\es\approx 17\text{eV}$ \rf{351n} for small values of $\zex\as (\ll 1)$ appear to be consistent with the numerical data.}
\end{table}

For an intuitive interpretation of the results of table~I it is important to first observe
that the relative derivation $\Delta$ of experimental data $\Delta\Eexp$ and electrostatic
RST predictions $\DERSTe$ (fourth column) increases from 1,7\% up to 11,5\% when
the nuclear charge~$\zex$ ranges from $z_{ex} = 32$ \kl germanium\kr\;up to $z_{ex}=83$ \kl
bismuth\kr. This effect can easily be understood in terms of the different relativistic
behaviour of the electric and magnetic interaction energies. Considering first the
electric type, one should recall that the {\it single\bs particle} energy $\ET$ \rf{267}
equals the mass eigenvalue $\Ms c^2$ \rf{325} which splits up into the matter energy $\ED$
\rf{246a} and external interaction energy $\Ees$ \rf{246c} according to
\begin{subequations}
\begin{align}
\label{345an}
\Ms c^2 &= Mc^2\sqrt{1 - \left(z_{ex}\as\right)^2} = \ED + \Ees \\
\label{345bn}
\ED &= \frac{Mc^2}{\sqrt{1 - \left(z_{ex}\as\right)^2}} \\
\label{345cn}
\Ees &= -\frac{\left(z_{ex}\as\right)^2}{\sqrt{1 - \left(z_{ex}\as\right)^2}}Mc^2 = \frac{1}{\sqrt{1 - \left(z_{ex}\as\right)^2}}\oEes \;,
\end{align}
\end{subequations}
\noindent see ref. \cite{Hu}. Here the non\bs relativistic limit $\oEes$ of the external interaction energy in the Coulomb field $\exAou$ \rf{39} is given by
\begin{eqnarray}
\label{346n}
\oEes &=& <\psi_0|{}^{(ex)}\!A_0|\psi_0> \\
\nonumber
 &=& \hbar c \vint\:\psi^{*}_0\vvr\frac{z_{ex}\as}{r}\psi\vvr = -\left(z_{ex}\as\right)^2 Mc^2 \;,
\end{eqnarray}
where the non\bs relativistic ground\bs state function $\psi_0\vvr$ of the single\bs particle Schr\"odinger problem \rf{330} coincides of course with the radial wave function $\oRp$ \rf{328}
\begin{equation}
\label{347n}
\psi_0\vvr = \frac{1}{\sqrt{4\pi}}\oRp = \sqrt{\frac{1}{\pi}\left(\frac{z_{ex}}{\aB}\right)^3}\exp\left[-\frac{z_{ex}r}{\aB}\right] \;.
\end{equation}
\indent From this result for the {\it external} interaction one may conclude that also the {\it internal} electrostatic interaction energy $\ERe$ \rf{258a} of both ground\bs state electrons is of the following form
\begin{equation}
\label{348n}
 \ERe = \frac{1}{\sqrt{1 - \left(z_{ex}\as\right)^2}}\oERe \approx \Delta\ERSTe 
\end{equation}
\noindent where $\oERe$ is again the non\bs relativistic limit \kl$z_{ex}\as\to 0$\kr\;of the internal electrostatic interaction energy $\ERe$. This non\bs relativistic limit may be simply determined by considering the electrostatic interaction energy of the two charge clouds due to the single\bs particle ground\bs state $\psi_0\vvr$ \rf{347n} of the ordinary Schr\"odinger problem \rf{330}:
\begin{eqnarray}
\label{349n}
\oERe &=& \iint {\rm d}^3\vec{r}\,{\rm d}^3\vec{r}\,'\:\frac{\left|\psi_0\vvr\right|^2\cdot\left|\psi_0\vvrs\right|^2}{\brrs} \\
\nonumber
 &=& \frac{5}{8}z_{ex}\frac{e^2}{\aB} \approx 17,00725\cdot z_{ex} \: \left[{\rm eV}\right] \;.
\end{eqnarray}
\noindent Combining this result with the former hypothesis \rf{348n} for the electrostatic interaction energy $\Delta\ERSTe$ says that the ``reference energy'' $\es$ defined through
\begin{equation}
\label{350n}
\es \doteqdot \frac{\sqrt{1 - \left(z_{ex}\as\right)^2}}{\zex}\cdot\Delta\ERSTe
\end{equation}
\noindent should adopt the value of \kl roughly\kr$17{\rm eV}$ for all values of nuclear
charge $\zex$. Now it is just this estimate which is confirmed by our numerical
calculations \kl table~I\kr\;within an error of a few percent when the elements from
$\zex=32$ \kl germanium\kr\; up to $\zex=83$ \kl bismuth\kr\;are considered. Indeed, the
value~$\varepsilon_0$ to be obtained by a combination of both equations~\rf{348n} and~\rf{349n}
\begin{equation}
\label{351n}
\varepsilon_0 = \frac{5}{8}\frac{e^2}{\aB} = 17,00725\ldots\left[{\rm eV}\right]
\end{equation}
\noindent appears as the upper limit for our RST calculations of $\es$ \rf{350n}, see table~I if the nuclear charge $\zex$ is adopted to become smaller \kl up to $\zex=2$ for the neutral helium; see a separate paper\kr. \\
\indent A similar estimate enlightens also the mechanism of the magnetic interactions, whose energy\bs momentum content is characterized by $\ERm$ \rf{259a}. In order to get the magnitude of the internal magnetic fields $\Hva$, we observe that when a point charge \kl emitting an electrostatic field $\vec{E}$ at rest\kr\;is moving with velocity $\vec{v}$, then there arises a magnetic field of magnitude
\begin{equation}
\label{352n}
\left|\vec{H}\right| \sim \frac{\frac{v}{c}}{\sqrt{1 - \left(\frac{v}{c}\right)^2}}\left|\vec{E}\right| \;.
\end{equation}
\noindent For an extended charge, one would include here some geometric factor $f_{*}$
being characteristic for the special charge distribution, and thus one would arrive at the
following relationship between the magnetic and electric field strengths
\begin{equation}
\label{353n}
\left|\Hv\right|^2 \sim f_{*}^2\frac{\left(\frac{v}{c}\right)^2}{1 - \left(\frac{v}{c}\right)^2} \cdot \left|\E\right|^2 \;.
\end{equation}
\noindent Now the kinetic energy part $\ED$ \rf{345bn} of the single\bs particle eigenvalue $\Ms c^2$ \rf{345an} suggests to take for the particle velocity in the ground\bs state
\begin{equation}
\label{354n}
\frac{v}{c} \sim \zex\as \;,
\end{equation}
\noindent and this yields the following relation between electric and magnetic energies $\Delta \ERSTe \approx \ERe$ \rf{258a} and $\DETmg$
\begin{equation}
\label{355n}
\DETmg \approx f_{*}^2\frac{\left(\zex\as\right)^2}{1 - \left(\zex\as\right)^2}\cdot \Delta\ERSTe \;.
\end{equation}
\indent Here it is very natural to assume that the magnetic interaction energy $\DETmg$ is responsible for the discrepancy of the experimental value $\Delta\Eexp$ and the electrostatic RST prediction $\Delta\ERSTe$, i.e. we rewrite equation \rf{355n} as
\begin{equation}
\label{356n}
\DETmg\equiv\Delta\Eexp - \Delta\ERSTe = f_{*}^2\frac{\left(\zex\as\right)^2}{1 - \left(\zex\as\right)^2}\cdot \Delta\ERSTe \;.
\end{equation}
\noindent This then leads us to the expectation that the geometric factor $\fs^2$
\begin{equation}
\label{357n}
\fs^2 = \frac{1 - \left(\zex\as\right)^2}{\left(\zex\as\right)^2}\cdot \frac{\Delta\Eexp - \Delta\ERSTe}{\Delta\ERSTe}
\end{equation}
\noindent will depend only very weakly , if at all, upon the coupling constant $\left(\zex\as\right)$. And indeed, if we insert the experimental values $\Delta\Eexp$ and our electrostatic RST results $\DERSTe$ into the right\bs hand side of equation \rf{357n}, one just finds the expected weak dependence of $\fs^2$, see for this table~I. \\
\indent Finally, combining both estimates for the electrostatic part $\DERSTe$ \rf{350n} and the magnetic part $\DETmg$ \rf{355n}, we arrive at the general form of the interaction energy $\Delta \Eexp$
\begin{eqnarray}
\label{358n}
\Delta\Eexp &=& \Delta\ERSTe + \DERSTmg \\
\nonumber
 &=& \frac{\zex}{\sqrt{1 - \left(\zex\as\right)^2}}\left\{1 + \fs^2\cdot\frac{\left(\zex\as\right)^2}{1 - \left(\zex\as\right)^2}\right\}\cdot\es \;.
\end{eqnarray}
\noindent This general result contains two slowly varying functions of the coupling constant $\left(\zex\as\right)$, which adopt their maximum values for small nuclear charge numbers \kl$\zex\to2$\kr. The maximum value of $\varepsilon_0$ of the reference energy $\es$ has already been determined in equation \rf{351n}, but for determining the maximal value \kl$f_0^2$, say\kr\;for the geometric factor $\fs^2$
\begin{equation}
\label{359n}
f_0^2 = 0,4
\end{equation}
\noindent one has to go deeply into the details of the magnetic interaction mechanism \kl see the deducation of equation \rf{643} below\kr. As we shall readily see, the crucial point here is the fact that the magnetic exchange forces do not vanish since there exists a non\bs trivial exchange current $\hv$, producing the magnetic exchange effects such as the magnetic exchange energy $\ECg$ \rf{259b}, whereas for the ground\bs state there is no {\it electric} exchange energy $\ECh$ \rf{258b} on behalf of the vanishing of the scalar exchange potential $B_0\vvr$!


\section{Magnetic Corrections}

The preceding estimate of the magnetic energy contributions hinted strongly upon a
magnetic explanation for the discrepancy between the experimental values $\Delta\Eexp$ and
the electrostatic RST predictions $\Delta\ERSTe$ \kl table~I\kr. Therefore it appears now
mandatory to explicitly compute the magnetic corrections, thought to be responsible for
the observed discrepancy. Through this procedure one will then obtain also more accurate
RST predictions. In the course of such an elaboration of the magnetic corrections it
should also become clear in which way the upper limit $f_0^2$ \rf{359n} of the slowly
varying function $\fs^2$ \rf{357n} comes about. Furthermore, the role of the magnetic
exchange effects , being induced by the ``magnetic'' exchange potential $\B$, has to be
clarified. Indeed, it will readily turn out that this ``magnetic'' exchange potential $\B$
cannot vanish for the reason of ground\bs state symmetry \kl i.e. isotropy\kr, which is in
sharp contrast to the missing of the ``electric'' exchange potential $\vec{B}_0\vvr$.
Actually, the ``magnetic'' exchange energy due to $\B$ will turn out to be {\it twice} the
magnetostatic contribution due to $\Ava$, which yields a higher
precision of the RST predictions by roughly one order of magnitude \kl compare the RST
results of tables~I and~II\kr. Consequently, the experimentally supported inclusion of the
``magnetic'' exchange effects due to $\Bv$ must be viewn as a strong confirmation of the \emph{non\bs abelian}  RST construction. Thus it becomes now necessary to work out the field theory of atomic magnetism in great detail. \\
\indent The desired magnetic corrections do emerge in two places: \\
\indent (i) as a small change $\Mamg$ of the mass eigenvalues $M_a$ \rf{34a}\bs\rf{34b}, \\
\indent (ii) as an additional contribution $\EGmg$ \rf{262} of the interaction energy $\EG$ \rf{260}. \\
\noindent Therefore the magnetic perturbation scheme must consist in first expressing the mass changes $\Mamg$ and energy change $\EGmg$ in terms of the wave functions $\psi_a(\vec{r})$ as solutions of the relativistic eigenvalue problem \rf{34a}\bs\rf{34b}, and then substituting  for these solutions their electrostatic approximations $\tilde{\psi}_a(\vec{r})$ as solutions of the truncated system \rf{36a}\bs\rf{36b}. In other words, one computes the value of the total energy functional $\ET$ upon the approximative solutions $\tilde{\psi}_a\vvr$ in place of the exact solutions $\psi_a\vvr$ of the original system \rf{34a}\bs\rf{34b}.

\subsection{Magnetic Mass Corrections $\Mamg$}

The point of departure for obtaining the ``magnetic'' mass corrections is the original exact form \rf{34a}\bs\rf{34b} of the two\bs particle problem. The corresponding eigenvalues $M_a$ are splitted into their electrostatic approximations $\tilde{M}_a$ and their magnetic corrections $\Mamg$
\begin{equation}
\label{41}
M_a = \tilde{M}_a + \Mamg.
\end{equation}
Quite generally, the exact mass eigenvalues $M_a$ can be expressed in terms of the corresponding eigenfunctions $\psi_a(\vec{r})$ by multiplying both sides of the eigenvalue equations \rf{34a}\bs\rf{34b} by $\bar{\psi}_a(\vec{r})$ and integrating, observing also the normalization relations \rf{335}. The magnetic mass corrections will then be due to those terms containing the vector potentials $\vec{A}_a\vvr = \{ {}^{(a)}\!A^{j}\vvr \}$ and $\vec{B}\vvr =\{ B^j\vvr\}$, which have been omitted for the electrostatic approximation. However, for discussing these magnetic terms it is instructive to reunite them with the electrostatic interaction terms and thus to consider the interelectronic interactions as a whole. This admits us to resort to the covariant Maxwell equations for carrying out the magnetic perturbation procedure where afterwards the separation of the electric and magnetic effects for obtaining the desired result $\Mamg$ will present no problem. \\
\indent In this sense, the united mass corrections $\Delta M_a$ may be deduced from the eigenvalue problem \rf{34a}\bs\rf{34b} in the following form:
\begin{subequations}
\begin{align}
\label{42a}
\hat{z}_1 \cdot \Delta M_1 c^2  &= -\hbar c \vint \left\{ \ekm \cdot \zAm + \hm \cdot \Bm \right\} \\
\label{42b}
\hat{z}_2 \cdot \Delta M_2 c^2  &= -\hbar c \vint \left\{ \zkm \cdot \eAm + \hms \cdot \Bms \right\} \; . 
\end{align}
\end{subequations} 
\noindent Obviously the mass corrections are built up by an electromagnetic part $\Maem$ and an exchange contribution $\Maxg$ 
\begin{equation}
\label{45}
\Delta M_a = \Maem + \Maxg \; ,
\end{equation}
\noindent with the self\bs evident arrangements 
\begin{subequations}
\begin{align}
\label{46a}
\hat{z}_1 \cdot \Meem c^2 &= - \hbar c \vint \;\; \ekm \cdot \zAm \\
\label{46b}
\hat{z}_2 \cdot \Mzem c^2 &= - \hbar c \vint \;\; \zkm \cdot \eAm \\
\label{46c}
\hat{z}_1 \cdot \Mexg c^2 &= - \hbar c \vint \;\; \hm \cdot \Bm \\
\label{46d}
\hat{z}_2 \cdot \Mzxg c^2 &= - \hbar c \vint \;\; \hms \cdot \Bms \; .
\end{align}
\end{subequations}
\noindent Here, the {\it electromagnetic} mass corrections $\Maem$ can be further split up according to
\begin{equation}   
\label{47}
\Maem = \Mae + \Mam
\end{equation}
\noindent with the electric part $\Mae$ being given by
\begin{subequations}
\begin{align}
\label{48a}
\hat{z}_1 \cdot \Mee c^2 &= - \hbar c \vint \;\; {}^{(1)}\!k_0\vvr \cdot \zAo \\
\label{48b}
\hat{z}_2 \cdot \Mze c^2 &= -\hbar c \vint \;\; {}^{(2)}\!k_0\vvr \cdot \eAo \; ,
\end{align}
\end{subequations}
\noindent and similarly for the magnetic part $\Mam$
\begin{subequations}
\begin{align}
\label{49a}
\hat{z}_1 \cdot \Mem c^2 &= - \hbar c \vint \;\; \ekj \cdot \zAj \equiv \hbar c \vint \;\; \vec{k}_1\vvr \cdot \vec{A}_2\vvr \\
\label{49b}
\hat{z}_2 \cdot \Mzm c^2 &= - \hbar c \vint \;\; \zkj \cdot \eAj \equiv \hbar c \vint \;\; \vec{k}_2\vvr \cdot \vec{A}_1\vvr \; .
\end{align}
\end{subequations}
\indent In a similar way, the exchange corrections $\Maxg$ \rf{46c}\bs \rf{46d} can also be subdivided into their ``electric'' parts $\Max$ and ``magnetic'' parts $\Mag$
\begin{equation}
\label{410}
\Maxg = \Max + \Mag \; ,
\end{equation}
\noindent i.e. we put
\begin{subequations}
\begin{align}
\label{411a}
\hat{z}_1 \cdot \Mex c^2 &= -\hbar c \vint \;\; \ho \cdot \Bo \\
\label{411b}
\hat{z}_2 \cdot \Mzx c^2 &= - \hbar c \vint \;\; \hos \cdot \Bos \\
\label{411c}
\hat{z}_1 \cdot \Meg c^2 &= -\hbar c \vint \;\; \hj \cdot \Bj \equiv \hbar c \vint \;\; \vec{h}\vvr \cdot \vec{B}\vvr \\
\label{411d}
\hat{z}_2 \cdot \Mzg c^2 &= - \hbar c \vint \;\; \hjs \cdot \Bjs \equiv \hbar c \vint \;\; \vec{h}^{*}\vvr \cdot \vec{B}^{*}\vvr \; .  
\end{align}
\end{subequations}
\indent The physical meaning of these mass corrections becomes now evident when they are
written in terms of the curvature components $F^{a}_{\;\;\mu \nu}$ and $G_{\mu \nu}$. Such
a reformulation of the mass corrections can easily be attained by eliminating the currents
$k_{a \mu}$ and $h_{\mu}$ from the original definitions \rf{46a}\bs \rf{46d} in favour of
the curvature components, namely just by means of the covariant Maxwell equations
\rf{236a}\bs\rf{236d}. Now, at this point we resort to an {\it additional} approximation
which considerably simplifies our magnetic perturbation approach, namely the
\emph{linearization} of just that gauge field dynamics \rf{236a}\bs\rf{236d}. This
linearization means concretely that \\
\begin{itemize}
\item[(i)] the cuvature components $F^a_{\;\;\mu \nu}$, $G_{\mu \nu}$ \rf{231a}\bs\rf{231e} become truncated into 
\begin{subequations}
\begin{align}
\label{412a}
F^a_{\;\;\mu \nu} &= \partial_{\mu} A^a_{\;\;\nu} - \partial_{\nu} A^a_{\;\;\mu} \\
\label{412b}
G_{\mu \nu} &= \partial_{\mu} B_{\nu} - \partial_{\nu} B_{\mu} 
\end{align}
\end{subequations}
by simply omitting the non\bs linear terms due to the non\bs abelian character of the original gauge group $U(2)$;
\end{itemize}
\noindent and
\begin{itemize}
\item[(ii)] the non\bs abelian Maxwell equations \rf{236a}\bs\rf{236d} become similarly truncated to their linear form
\begin{subequations}
\begin{align}
\label{413a}
\partial^{\mu} F^a_{\;\;\mu \nu} &= - 4 \pi \alpha_{\txt{s}} \:k^a_{\;\;\nu} \\
\label{413b}
\partial^{\mu} G_{\mu \nu} &= 4 \pi \alpha_{\txt{s}} \: h^{*}_{\nu} \; .
\end{align}
\end{subequations}
\end{itemize}

\indent As a consequence of these simplifying assumptions, one finds for those four\bs vector products dertermining the mass corrections $\Maem$ and $\Maxg$ \rf{46a}\bs \rf{46d}: 
\begin{subequations}
\begin{align}
\label{414a}
\ekm \cdot \zAm &= - \frac{1}{4 \pi \as} \partial^{\mu} \Big\{ \eFmnu \cdot \zAn \Big\} + \frac{1}{8\pi\as} \eFmnu \cdot \zFmno \\
\label{414b}
\zkm \cdot \eAm &= - \frac{1}{4\pi\as} \partial^{\mu} \Big\{ \zFmnu \cdot \eAn \Big\} + \frac{1}{8\pi\as} \zFmnu \cdot \eFmno \\
\label{414c}
\hm \cdot \Bm &= \frac{1}{4\pi\as} \partial^{\mu} \Big\{ \Gmnus \cdot \Bn \Big\} - \frac{1}{8\pi\as} \Gmnos \cdot \Gmnu \;.
\end{align}
\end{subequations}
\noindent When this is substituted back into the correction formulae \rf{46a}\bs \rf{46d}, they appear in a new form by means of Gauss' integral theorem, namely as
\begin{subequations}
\begin{align}
\label{415a}
\hat{z}_1 \cdot \Meem c^2 &= \hat{z}_2 \cdot \Mzem c^2 = - \frac{\hbar c}{8\pi\as} \vint \;\; \eFmnu \cdot \zFmno \\
\nonumber
 & \\
\label{415b}
\hat{z}_1 \cdot \Mexg c^2 &= \hat{z}_2 \cdot \Mzxg c^2 = \frac{\hbar c}{8\pi\as} \vint \;\; \Gmnos \cdot \Gmnu \; .
\end{align}
\end{subequations}
\noindent This result explicitly demonstrates that the mass corrections $\Delta M_a$ \rf{42a}\bs \rf{42b} for both particles \kl$a = 1,2$\kr are actually identical \kl$\hat{z}_1 \cdot \Delta M_1 \equiv \hat{z}_2 \cdot \Delta M_2$\kr . \\
\indent But once that ``Lorentz invariant'' form \rf{415a}\bs \rf{415b} of the mass corrections is known, it is self\bs suggesting to split them up into two contributions, which are offered by themselves through the introduction of the electrostatic and magnetostatic field strengths $\Ea = \left\{ \aEj \right\}$ and $ \Ha = \left\{ \aHj \right\}$, i.e. we put
\begin{subequations}
\begin{align}
\label{416a}
\aEj &\doteqdot {}^{(a)}\!F_{0 j}\vvr = - \partial_j {}^{(a)}\!A_0\vvr \\
\label{416b}
\aHj &\doteqdot \frac{1}{2} \varepsilon^{j k}_{\;\;\;\; l} \;{}^{(a)}\!F_k^{\;\;l}\vvr = \varepsilon^{j k}_{\;\;\;\; l} \: \partial_k {}^{(a)}\!A^l\vvr \; , 
\end{align}
\end{subequations}
\noindent or in three\bs vector notation
\begin{subequations}
\begin{align}
\label{417a}
\Ea &= - \vec{\nabla} {}^{(a)}\!A_0\vvr \\
\label{417b}
\Ha &= \vec{\nabla} \times \vec{A}_a\vvr \; .
\end{align}
\end{subequations}    
\indent Now by this arrangement, the Lorentz invariant product of curvature components determining the electromagnetic mass corrections $\Maem$ \rf{415a} reads in three\bs vector notation 
\begin{equation}
\label{418}
\eFmnu \cdot \zFmno = 2 \left[ \He \cdot \Hz - \Ee \cdot \Ez \right] \; ,
\end{equation}
\noindent and clearly this yields a natural splitting of those electromagnetic mass corrections \rf{415a}, namely just the former equation \rf{47} with the following identifications for the electric part
\begin{equation}
\label{419}
\hat{z}_1 \cdot \Mee c^2 = \hat{z}_2 \cdot \Mze c^2 = \frac{\hbar c}{4 \pi \as} \vint \;\; \Ee \cdot \Ez \; , 
\end{equation}
\noindent and similarly for the magnetic part
\begin{eqnarray}
\label{420}
\hat{z}_1 \cdot \Mem c^2 = \hat{z}_2 \cdot \Mzm c^2 = - \frac{\hbar c}{4\pi\as} \vint \;\; \He \cdot \Hz \; .
\end{eqnarray}
\noindent Observe here the curious fact that the magnetic part \rf{420} enters the electromagnetic mass correction $\Maem$ \rf{47} with the {\it opposite} sign in comparison to the electric part \rf{419} which is a consequence of the Lorentz invariance of the product of the field strenghts \rf{418}! \\
\indent Clearly, it is self\bs suggesting now to treat the exchange corrections $\Maxg$ \rf{415b} in a quite similar way. This means that one introduces an ``electric'' exchange vector field $\X = \left\{ X^j \vvr \right\}$ \kl as the exchange counterpart of $\vec{E}\vvr$ \rf{416a}\kr and also a ``magnetic'' exchange field $\Y = \left\{ Y^j \vvr \right\}$ \kl as the exchange counterpart of $\vec{H}\vvr$ \rf{416b}\kr through
\begin{subequations}
\begin{align}
\label{421a}
X^j \vvr &\doteqdot G_{0 j} \vvr = - \partial_j B_0 \vvr - \frac{i}{\aM} B_j \vvr \\
\label{421b}
Y^j \vvr &\doteqdot \frac{1}{2} \varepsilon^{jkl} G_{kl} \vvr = \varepsilon^{jk}_{\;\,\;\;l} \partial_k B^l \vvr \; ,
\end{align}
\end{subequations}
\noindent where the {\it exchange length parameter} $\aM$ is given by
\begin{equation}
\label{422}
\aM \doteqdot \frac{\hbar}{(M_1 - M_2) c^2} \; .
\end{equation}
\noindent In three\bs vector notation, the relations \rf{421a}\bs \rf{421b} read
\begin{subequations}
\begin{align}
\label{423a}
\X &= - \vec{\nabla} B_0\vvr + \frac{i}{\aM} \B \\
\label{423b}
\Y &= \vec{\nabla} \times \B \; ,
\end{align}
\end{subequations}
\noindent and thus  obviously represent the exchange analogue of the corresponding electromagnetic relations \rf{417a}\bs \rf{417b}, however with the difference that there emerges now a typical length parameter $\aM$ \rf{422} which gives an inherent measure for the spatial range of the exchange effects. In terms of these new exchange fields, the Lorentz scalar for the exchange mass corrections $\Maxg$ \rf{415b} reads
\begin{equation}
\label{424}
\Gmnos \cdot \Gmnu = 2 \left[ \Y^{*} \cdot \Y - \X^{*} \cdot \X \right] \; ,
\end{equation}
\noindent which is of course again the exchange analogue of the corresponding electromagnetic relation \rf{418}. As a result, the exchange mass corrections $\Maxg$ \rf{415b} can ultimately be written as
\begin{equation}
\label{425}
\hat{z}_1 \cdot \Mexg c^2 = \hat{z}_2 \cdot \Mzxg c^2 = \frac{\hbar c}{4\pi\as} \vint \left[ \Y^{*} \cdot \Y - \X^{*} \cdot \X \right] \; .
\end{equation}
\indent Though this result looks formally quite analogous to the electromagnetic case, one nevertheless cannot relate here the ``electric'' exchange masses $\Max$ \rf{411a}\bs \rf{411b} to the ``electric'' exchange vector $\X$ and analogously the ``magnetic'' corrections $\Mag$ \rf{411c}\bs \rf{411d} to the ``magnetic'' exchange vector $\Y$ as it was done for the electromagnetic situation \rf{419}\bs \rf{420}. Instead, the correct relationships for the ``electric'' exchange mass must look as follows:
\begin{equation}
\label{426}
\hat{z}_1 \cdot \Mex c^2 = \hat{z}_2 \cdot \Mzx c^2 = - \frac{\hbar c}{4\pi\as} \vint \;\; \X^{*} \cdot \X + \frac{\hbar c}{4\pi\as} \frac{1}{\aM^2} \vint \;\; \B^{*} \cdot \B \; .
\end{equation}
\noindent Thus both ``electric'' exchange corrections are again identical, as it is the
case also for the electromagnetic subsystem \rf{419}. However, they do not coincide with the ``electric'' part due to $\X$ of the total exchange corrections $\Maxg$ \rf{425}. \\
\indent A similar effect occurs also with the ``magnetic'' parts $\Mag$ of the exchange mass corrections $\Maxg$ \rf{425}. Starting from the original definitions \rf{411c}\bs \rf{411d} one finds again that both ``magnetic'' contributions are identical 
\begin{equation}
\label{427}
\hat{z}_1 \cdot \Meg c^2 = \hat{z}_2 \cdot \Mzg c^2 = \frac{\hbar c}{4\pi\as} \vint \;\; \Y^{*} \cdot \Y - \frac{\hbar c}{4\pi\as} \frac{1}{\aM^2} \vint \; \vec{B}^{*}\vvr \cdot \B \; ,
\end{equation}
\noindent but they do again not coincide with the ``magnetic'' part \kl due to $\Y$\kr of the total exchange masses $\Maxg$ \rf{425}. However, if both the ``electric'' and the ``magnetic'', contributions \rf{426} and \rf{427} are added up separately for either particle \kl $a = 1,2$\kr, one just finds the total exchange corrections $\Maxg$ \rf{425}. Thus the electromagnetic and exchange subsystems are found to differ in the way in which the total corrections are distributed upon the ``electric'' and the ``magnetic'' subsystems.

\subsection{Linearized Gauge Field Equations}

Obviously, the introduction of the three\bs vector notation enables one to separate uniquely the electric corrections \kl being already included in the electrostatic approximation\kr\;from the magnetic corrections which will be treated subsequently by use of an adequate perturbation approach. Consequently, one would like to write down now the gauge field equations \rf{236a}\bs\rf{236d} in three\bs vector notation where one simultaneously restricts oneself to their linearized form \rf{413a}\bs\rf{413b}. This implies that the magnetic corrections will be taken into account only in their lowest\bs order approximation. However, the consistent linearization of the gauge field equations represents a certain problem which requires now an extra discussion. \\
\indent First consider the electromagnetic subsystem, for which one concludes from the linearized Maxwell equations \rf{413a} that the three\bs current $\kva = \left\{ {}^{(a)}\!k^{j}(\vec{r}) \right\}$ must have vanishing divergence \kl$a=1,2$\kr
\begin{equation}
\label{428}
\nv \cdot \kva = 0 \; .
\end{equation}
\noindent Thus the linearization of the \kl non\bs Abelian\kr Maxwell equations \rf{236a}\bs\rf{236b} must imply the neglection of the right\bs hand sides of the source equations \rf{240a}\bs\rf{240b}. Furthermore, when the curvature components $F^a_{\;\;\mu \nu}$ are expressed through the connection components $A^a_{\;\;\mu}$ \rf{412a}, that equation \rf{413a} yields the corresponding Poisson equations
\begin{subequations}
\begin{align}
\label{429a}
\Delta \aAo &= 4\pi\as \cdot \ako \\
\label{429b}
\Delta \Ava &= 4\pi\as \cdot \kva \; ,
\end{align}
\end{subequations}
\noindent provided one subjects the magnetic vector potentials $\Ava$ \kl$a= 1,2$\kr to the Coulomb gauge condition
\begin{equation}
\label{430}
\nv \cdot \Ava = 0 \; .
\end{equation}
\noindent Observing here the usual boundary conditions at infinity \kl$r \to \infty$\kr , the solutions of the Poisson equations \rf{429a}\bs \rf{429b} are easily found as
\begin{subequations}
\begin{align}
\label{431a}
\aAo &= -\as \vints \;\; \frac{\akos}{| \vec{r} - \vec{r'}|} \\
\label{431b}
\Ava &= -\as \vints \;\; \frac{\kvas}{| \vec{r} - \vec{r'}|} \; ,
\end{align}
\end{subequations}
\noindent where the spatial part \rf{431b} is easily seen to meet with the gauge condition \rf{430}. Thus it is evident that the linearized electromagnetic subsystem is governed by the usual \kl Abelian\kr Maxwellian structure, which may be expressed also through the field equations for the field strenghts $\Eva$, $\Hva$ \rf{417a}\bs \rf{417b}:
\begin{subequations}
\begin{align}
\label{432a}
\nv \cdot \Eva &= - 4\pi \as \cdot \ako \\
\label{432b}
\nv \cdot \Hva &= 0 \\
\label{432c}
\nv \times \Eva &= 0 \\
\label{432d}
\nv \times \Hva &= -4\pi\as \cdot \kva \; .
\end{align}
\end{subequations}
\indent However, the exchange subsystem has a somewhat different structure. This became already obvious through the introduction of the ``electric'' exchange field $\X$ \rf{423a} which is not simply the gradient of the ``electric'' exchange potential $B_0 \vvr$ but contains also the ``magnetic'' exchange potential  $\Bv$! Nevertheless, one finds from the linearized equations \rf{413b} again the ordinary Poisson equation for $B_0\vvr$
\begin{subequations}
\begin{align}
\label{433a}
\Delta B_0 \vvr &= -4\pi\as \hos \\
\label{433b}
B_0 \vvr &= \as \vints \frac{\hoss}{\brrs} \; ,
\end{align}
\end{subequations}
\noindent which is thus revealed as not being affected by taking into account the magnetic corrections \kl similar as for the electric potentials $\aAo$ \rf{429a}\kr. Observe also, that for the deduction of the exchange Poisson equation \rf{433a} we imposed a Coulomb\bs like gauge condition upon the ``magnetic'' exchange potential $\Bv$
\begin{equation}
\label{434}
\nv \cdot \Bv = 0 \; ,
\end{equation}
\noindent in close analogy to the magnetic counterpart $\Ava$ \rf{430}. Strictly speaking, such gauge fixing conditions as \rf{430} must not be imposed upon the fields $B_{\mu}$ because they lost their geometric status as gauge potentials in the process of Abelian symmetry breaking and therefore obey a {\it homogeneous} transformation law \cite{Hu}. However, it is easy to see that the requirement \rf{434} is consistent with all the other static field equations for the exchange variables $B_0\vvr$, $\Bv$. \\
\indent  It is true, the ``electric'' part of the exchange field equations \rf{433a}\bs \rf{433b} is not too much different from its electric counterpart \rf{429a}and \rf{431a}; however, for the exchange vector potential $\Bv$ one finds from the linear equations \rf{413b}
\begin{equation}
\label{435}
\Delta \Bv + \frac{1}{\aM^2} \cdot \Bv = - 4\pi\as \hvs - \frac{i}{\aM} \nv B_0\vvr \; .
\end{equation}
\noindent As a consistency test of this equation one forms here the divergence of the left- and right\bs hand side and arrives just at the Poisson equation \rf{433a} for the exchange potential $B_0\vvr$, provided the divergence relation for $\Bv$ \rf{434} is respected together with the following source equation for the exchange current $\hv$:
\begin{equation}
\label{436}
\nv \cdot \hvs = \frac{i}{\aM} \hos \; .
\end{equation}
\noindent However, this is nothing else than the continuity equation for the exchange four\bs current $h_{\mu}^{*}$
\begin{equation}
\label{437}
\partial^{\mu} \hms = 0
\end{equation}
\noindent which itself is a consequence of the {\it linear} exchange field equations \rf{413b} and, on the other hand, coincides with the former source equations \rf{240c}\bs\rf{240d} if the right\bs hand sides vanish \kl to be justified below\kr. In this way, one actually attains a consistent linear approximation of the non\bs Abelian \kl and therefore non\bs linear\kr\, gauge field equations. \\
\indent Of course, the solutions $\Bv$ of the exchange field equation \rf{435} must have a somewhat other shape than their magnetic counterparts $\Ava$ \rf{431b}:
\begin{equation}
\label{438}
\Bv = \as \vints \frac{1}{\brrs} \cdot \cos\left( \frac{\brrs}{a_{\rm{M}}} \right) \cdot \left\{ \hvsp + \frac{i}{4\pi\as\aM} \cdot \nvs B_0\vvrs \right\} \; .
\end{equation}
\noindent Thus the effect of the exchange length $\aM$ is to suppress the magnetic exchange vector potential $\Bv$ if the exchange current $\vec{h}\vvr$ is smoothly spread over a spatial domain much larger than the exchange length $\aM$. But if $\vec{h}\vvr$ is well\bs localized within such an ``exchange domain'', the vector potential $\Bv$ is non\bs zero, but fades away as $r^{-1} \cdot \cos\left( \frac{r}{\aM} \right)$. On the other hand, for $\aM \to \infty$ the ``magnetic'' exchange equation \rf{435} degenerates to an ordinary Poisson equation of the kind \rf{433a} or \rf{429a}\bs \rf{429b} with the corresponding behaviour of the solutions. Thus the ``magnetic'' exchange effects are found to be of a rather different type which may again be expressed more concisely by the following ``exchange equations'':
\begin{subequations}
\begin{align}
\label{439a}
\nv \cdot \X &= 4\pi\as \hos \\
\label{439b}
\nv \times \X &= \frac{i}{\aM} \Y \\
\label{439c}
\nv \cdot \Y &= 0 \\
\label{439d}
\nv \times \Y &= 4\pi\as \cdot \hvs + \frac{1}{\aM^2} \cdot \Bv + \frac{i}{\aM} \cdot \nv B_0\vvr \; ,
\end{align}
\end{subequations}
\noindent which adopt the ordinary Maxwellian form \rf{432a}\bs \rf{432d} for infinite exchange length \kl$\aM \to \infty$\kr. As a consistency test for the linearization procedure, apply the divergence operation to the last equation \rf{439d} and find just the source equation for the exchange current $\hv$ \rf{436} by use of the Poisson equation \rf{433a}.

\subsection{Mass Corrections and Densities}

Once both the electromagnetic potentials $\left\{ \aAo \:;\: \Ava \right\}$ and the exchange potentials $\left\{ B_0\vvr \:;\: \Bv \right\}$ are known in terms of the charge and current densities, it becomes possible to eliminate these potentials completely from the mass corrections and to express the latter objects exclusively in terms of these physical densities. Clearly, this then represents a considerable technical simplification because one can express the desired mass corrections directly in terms of the wave functions whose link to the current densities is well\bs known, see equations \rf{38a}\bs\rf{38b}. Thus to begin with, reconsider the electric corrections $\Mae$ \rf{48a}\bs \rf{48b} and substitute therein for the electrostatic potentials $\aAo$ their general form \rf{431a} in order to arrive at
\begin{equation}
\label{440}
\hat{z}_1 \cdot \Mee c^2 = \hat{z}_2 \cdot \Mze c^2 = e^2  \iint d^3\vec{r}\, d^3\vec{r}\,' \frac{\eko \cdot \zkop}{\brrs} \;.
\end{equation}
\noindent Of course, the charge densities $\ako$ may here be further expressed in terms of the Dirac wave functions $\psi_a\vvr$ according to the former relations \rf{38a}. Observe also, that the present result of the electrostatic interaction energy $\hat{z}_a \cdot \Mae c^2$ now admits two different interpretations, namely either as an instantaneous Coulomb interaction between the two charge clouds $\ako$ or as the {\it interaction} energy \kl {\it not} self\bs energy!\kr of the electric field modes $\Eva$ emitted by the charge clouds, see equation \rf{419}. This bilinear \kl instead of quadratic\kr\;construction for the electromagnetic interaction energy of the particles has a certain tradition in the literature and emerges also quite naturally in the RST formalism \cite{Ma2}.\\
\indent A similar effect is obtained also for the magnetic corrections $\Mam$ \rf{49a}\bs \rf{49b} which by means of the vector potentials $\Ava$ \rf{431b} may be recast into the following form
\begin{equation}
\label{441}
\hat{z}_1 \cdot \Mem c^2 = \hat{z}_2 \cdot \Mzm c^2 = -e^2 \iint d^3\vec{r}\, d^3\vec{r}\,' \frac{\kve \cdot \kvzs}{\brrs} \; .
\end{equation}
\noindent Thus the magnetostatic interaction energy $\hat{z}_a \cdot \Mam c^2$ can also be interpreted as being due to either an instantaneous direct interaction of the currents $\kva$ or as the interaction energy of the magnetic field modes $\Hva$ \rf{420}. It is also interesting to remark, that the the sum of the electric and magnetic corrections appears as the integral of a four\bs vector product, namely by adding both equations \rf{440} and \rf{441}:
\begin{eqnarray}
\label{442}
\hat{z}_a \cdot \Mae c^2 + \hat{z}_a \cdot \Mam c^2 &\equiv& \hat{z}_a \cdot \Maem c^2 \\
\nonumber
 &=& e^2 \iint d^3\vec{r}\, d^3\vec{r}\,' \frac{\akm \cdot \akmos}{\brrs} \\
\nonumber
\big( a &=& 1\;\text{or}\;2 \big) \; .
\end{eqnarray}

Now it seems natural to expect that the exchange field system will display an analogous structure. This however is true only for its ``electric'' component, but not for the ``magnetic'' one. In order to see this more clearly, first recall that the ``electric'' exchange potential $B_0\vvr$ \rf{433a}\bs \rf{433b} is of the same structure as its electric counterpart $\aAo$ \rf{431a} and therefore yields for the ``electric'' exchange mass $\Max$ \rf{411a}\bs \rf{411b}
\begin{equation}
\label{443}
\hat{z}_1 \cdot \Mex c^2 = \hat{z}_2 \cdot \Mzx c^2 = -e^2 \iint d^3\vec{r}\, d^3\vec{r}\,' \frac{\ho \cdot \hoss}{\brrs} \; .
\end{equation}
\noindent Indeed, this looks very similar to the electrostatic corrections $\Mae$ \rf{440} where the charge densities $\ako$ play the part of the exchange densities $\ho$, $\hos$ and a change of sign does occur additionally. This {\it lowering} of the electrostatic interaction energy \rf{440} by the exchange energy \rf{443} is due to the fact that the present RST formalism is equivalent to the {\it antisymmetrized} product states of the conventional Hartree\bs Fock approach. \\
\indent However, the ``magnetic'' exchange system appears in a somewhat different shape which traces back to the modified Poisson equation \rf{435} for the ``magnetic'' exchange potential $\Bv$. More concretly, introducing the solution \rf{438} for $\Bv$ into the exchange correction formulae \rf{411c}\bs \rf{411d} yields
\begin{subequations}
\begin{align}
\label{444a}
\hat{z}_1 \cdot \Meg c^2 = \: &e^2 \iint d^3\vec{r}\, d^3\vec{r}\,' \frac{ \cos \left( \frac{\brrs}{\aM} \right) \hv \cdot \hvsp}{\brrs} \\
\nonumber
 &+ \frac{i \hbar c}{4\pi\aM} \iint d^3\vec{r}\, d^3\vec{r}\,' \frac{ \cos \left( \frac{\brrs}{\aM} \right) \hv \cdot \nvs B_0\vvrs}{\brrs} \\
\nonumber
 & \\
\label{444b}
\hat{z}_2 \cdot \Mzg c^2 = \: &e^2 \iint d^3\vec{r}\, d^3\vec{r}\,' \frac{ \cos \left( \frac{\brrs}{\aM} \right) \hv \cdot \hvsp}{\brrs} \\
\nonumber
 &- \frac{i \hbar c}{4\pi\aM} \iint d^3\vec{r}\, d^3\vec{r}\,' \frac{ \cos \left( \frac{\brrs}{\aM} \right) \hvs \cdot \nvs B_0^{*}\vvrs}{\brrs} \;.
\end{align}
\end{subequations}
\noindent Obviously, the presence of the terms containing the gradient of the ``electric'' exchange potential $B_0\vvr$ does not allow us to recast this result again into the form of an integral over the four\bs vector product $\hms \cdot h^{\mu}\vvr$, as it was the case for the electromagnetic analogue \rf{442}. \\
\indent After the energy contributions of the electric and magnetic type have been discussed in detail, one can now render more precise the meaning of the {\it ``magnetic corrections''}. Obviously, the purely electric corrections $\Maex$ \kl to be included into the electrostatic approximation\kr\;are given by all the mass corrections of the ``electric'' type \rf{440} and \rf{443}
\begin{equation}
\label{445}
\Maex \doteqdot \Mae + \Max \; ,
\end{equation}
\noindent whereas the magnetic corrections \kl to be superimposed as perturbations over the results of the electrostatic approximation\kr are given by the correction terms of the ``magnetic'' type \rf{441} and \rf{444a}\bs \rf{444b}\,:
\begin{equation}
\label{446}
\Mamg \doteqdot \Mam + \Mag \; .
\end{equation}
\noindent The lowest\bs order approximation of such a perturbation scheme consists then in inserting the solutions $\tilde{\psi}_a\vvr$ of the electrostatic approximation system \rf{36a}\bs\rf{36b} into these magnetic correction functionals \rf{446}.

\subsection{Magnetic Energy of the Gauge Field}

Besides the magnetic contributions $\Mamg$ as part of the mass eigenvalues $M_a$, there occur  further magnetic constituents of the total energy $\ET$, namely those being due to the gauge field energy $\EG$ \rf{253}. Since the gauge field energy $\EG$ \rf{253} itself splits up into the electromagnetic part $\ER$ \rf{254} and exchange part $\EC$ \rf{255}, one may separate again both parts into the electric and magnetic type \rf{257a}\bs\rf{257b}. Here the electric part $\ERe$ \rf{258a} is naturally included in the electrostatic approximation and is found now to agree just with the electrostatic mass corrections $\Mae$ \rf{419}:
\begin{equation}
\label{448}
\ERe \equiv \hat{z}_a \cdot \Mae c^2 \;\; , \;\; \kl a = 1 \mbox{\;or\;} 2\kr.
\end{equation}
\noindent This is the reason why the {\it subtraction} of $\EGeh$ from the sum of mass eigenvalues rescinds the double counting of the electrostatic energy $\ERe$ for the total energy $\ET$ \rf{339}. Similarly, the magnetic part $\ERm$ of the electromagnetic gauge field energy $\ER$ \rf{257a} reads in terms of the magnetic field strenghts $\Hva$
\begin{equation}
\label{449}
\ERm = \frac{\hbar c}{4\pi\as} \vint \; \He \cdot \Hz 
\end{equation}
\noindent and thus is the {\it negative} of the magnetostatic mass corrections $\Mam$ \rf{420}
\begin{equation}
\label{450}
\ERm = - \hat{z}_a \cdot \Mam c^2 \;\;\;\;\kl a = 1 \mbox{\;or\;} 2\kr \; ,
\end{equation}
\noindent in contrast to the corresponding electric counterpart \rf{448}. Therefore the {\it addition} of $\EGmg$ to the sum of mass eigenvalues rescinds the double counting of the magnetostatic energy $\ERm$ for the total energy $\ET$ \rf{339}. \\
\indent In an analoguous way, the exchange energy $\EC$ \rf{257b} may also be split up into the contributions of ``electric'' and ``magnetic'' type according to $\ECh$ \rf{258b} and $\ECg$ \rf{259b}. The ``electric'' part $\ECx$ however cannot be identified here with the mass corrections $\Max$ \rf{411a}\bs\rf{411b}, as it was possible for the electrostatic approximation, but the exchange vector potential $\Bv$ has to be retained so that one first finds 
\begin{equation}
\label{452}
\ECx = \frac{\hbar c}{4\pi\as\aM^2} \vint \; \Bvs \cdot \Bv + \frac{\hbar c}{4\pi\as} \vint \; \nv B_0^{*}\vvr \cdot \nv B_0^{*}\vvr \; .
\end{equation}
\noindent Clearly, when the exchange vector potential $\Bv$ is neglected here, one returns to the truncated form due to the electrostatic approximation. On the other hand, eliminating the ``electric'' exchange potential $B_0\vvr$ from the present result for $\ECx$ \rf{452} in favour of $\X$ by means of its definition \rf{423a} yields now the generalization of the former electrostatic approximation, namely for both $a = 1$ or $a = 2$:
\begin{eqnarray}
\label{453}
\ECx &=& \frac{\hbar c}{4\pi\as} \vint \; \Xs \cdot \X \\
\nonumber
 &=& - \hat{z}_a \cdot \Max c^2 + \frac{\hbar c}{4\pi\as\aM^2} \vint \; \Bvs \cdot \Bv\; .
\end{eqnarray}
Thus it is seen that the ``electric'' exchange energy $\ECx$ receives an additional contribution in comparison to its mass correction counterpart  $\Max$ \rf{411a}\bs\rf{411b} when the ``magnetic'' exchange potential $\Bv$ is not zero. This additional contribution \kl i.e. the integral on the right\bs hand side of \rf{453}\kr\;must be taken into account for building up the ``magnetic'' energy correction $\Delta \EGmg$ below; whereas the first contribution, i.e. the ``electric'' exchange mass $\Max$ \rf{443}, is already contained in the electrostatic approximation of $\tilde{E}_T$ \rf{342} and must therefore be omitted for the magnetic part of $\Delta \EGmg$. \\
\indent In a quite analogous manner one can treat the ``magnetic'' exchange energy $\ECg$ \rf{259b} which is nothing else than the exchange analogue of the magnetostatic field energy $\ERm$ \rf{449}. Comparing this again to the ``magnetic'' exchange masses $\Mag$ \rf{427} yields the relationship
\begin{equation}
\label{455}
\ECg = \hat{z}_a \cdot \Mag c^2 + \frac{\hbar c}{4\pi\as\aM^2} \vint \; \Bvs \cdot \Bv \; .
\end{equation}
\noindent Here it is interesting to observe that the energy corrections of the {\it electric} type, namely $\ERe$ \rf{448} and $\ECx$ \rf{453}, contain the corresponding mass corrections $\Mae$ and $\Max$ with opposite sign so that they actually become subtracted from the sum of mass eigenvalues $M_a$ in order to build up the total energy $\ET$ \rf{339}. As a result, the double counting of these corrections of the ``electric'' type is actually avoided, as explained in connection with the electrostatic approximation \kl see the discussion below equation \rf{341}. \\
\indent However, in contrast to this, the corrections of the magnetic type, i.e. $\ERm$ \rf{450} and $\ECg$ \rf{455} both contain the ``magnetic'' mass corrections $\Mam$ and $\Mag$ with the ``false'' sign; but since $\EGmg$ \rf{262} has to be {\it added} to the same terms occuring in the sum of mass corrections $M_a$ for building up the total energy $\ET$ \rf{339} the double counting of the ``magnetic'' energy contributions is rescinded in just the same way as it is the case with the ``electric'' contributions. For the magnetic energy correction $\Delta \EG^{(mg)}$ we therefore find
\begin{eqnarray}
\label{456}
\Delta \EG^{(mg)} &=& \ERm - \ECg - \frac{\hbar c}{4\pi\as} \vint \; \Bvs \cdot \Bv \\
\nonumber
 &=& - \hat{z}_a \cdot \Mam c^2 - \hat{z}_a \cdot \Mag c^2 - \frac{\hbar c}{2\pi\as\aM^2} \vint \; \Bvs \cdot \Bv \; .
\end{eqnarray}
\noindent Observe here that half of the integral on the right\bs hand side emerges for both the ``electric'' part $\ECx$ \rf{453} and the magnetic part $\ECg$ \rf{455} and thus must enter the energy correction $\Delta \EG^{(mg)}$ \rf{456} with its double value. \\
\indent Summarizing, the total energy correction $\DETmg$ of the ``magnetic'' type is now found as the sum of all those terms in the total energy $\ET$ \rf{339} which were omitted for the electrostatic approximation $\tilde{E}_T$ \rf{342}, i.e.
\begin{eqnarray}
\nonumber
\DETmg &\doteqdot& \sum_{a=1}^{2} \hat{z}_a M_a^{(mg)} c^2 + \Delta \EG^{(mg)} \\
\label{457}
 &=& \frac{1}{2} \sum_{a=1}^{2} \hat{z}_a  M_a^{(mg)} c^2 -  \frac{\hbar c}{2\pi\as\aM^2} \vint \Bvs \cdot \Bv \; .
\end{eqnarray}
\noindent Here the ``magnetic'' masses $M_a^{(mg)}$ are the proper sum of the
magnetostatic mass $\Mam$ \rf{441} and its exchange counterpart $\Mag$ \rf{444a}\bs
\rf{444b}, see equation \rf{446}. Clearly, the expectation is now that, when the magnetic
energy correction $\DETmg$ \rf{457} is added to the electrostatic approximation
$\tilde{E}_T$ \rf{342} one will arrive at a more precise numerical prediction for the
atomic energy levels $\ET$. However, before this expectation receives its validation, it
is very instructive to convince oneself of the physical correctness of the RST picture of the atomic magnetism by considering the interaction with an {\it external} magnetic field \kl Zeeman effects\kr.


\section{External Magnetism}

It may seem somewhat strange that the magnetic contributions $\EGmg$ are entering the total energy $\ET$ \rf{339} with the opposite sign in comparison to the electric contributions $\Delta\EGeh$. This is the more amazing as the magnetostatic energy content $\ERm$ \rf{259a} is also given by the {\it positive} product of the magnetostatic fields $\Ha$, quite analogously to the electrostatic counterpart $\ERe$ \rf{258a}. However, the formalism of minimal coupling \rf{35a}\bs\rf{35b} together with Lorentz invariance inevitably leads to that minus sign for the magnetostatic corrections as part of the total energy $\ET$; and therefore one may wish to have some additional supporting argument that the magnetic interactions are actually implemented in the RST dynamics in the right way. Such an additional argument can be put forward by considering the interaction of either of the two electrons with an external constant magnetic field $\vec{H}_{ex}$. Indeed, the covariant derivatives \rf{35a}\bs\rf{35b} most clearly display the fact that the corresponding vector potential $\vec{A}_{ex}\vvr$
\begin{equation}
\label{525}
\nv \times \vec{A}_{ex}\vvr = \vec{H}_{ex}
\end{equation}
\noindent acts upon the $a$\bs th particle in just the same way as does the vector
potential ${}^{(b)}\!\vec{A}\vvr \; (b \neq a)$ due to the other particle. Therefore, we
can test the correctness of the magnetostatic interelectronic interactions by simply
inspecting the interaction with an external source emitting the constant field
$\vec{H}_{ex}$ \kl Zeeman effect, see e.g.~\cite{Me2}\kr. \\
\indent Now it is well well\ known that the interaction of a bound electronic system with an external magnetic field is phenomenologically described by including into the \kl non\bs relativistic\kr\: Hamiltonian $\hat{H}$ an interaction term $\hat{H}_{int}$ of the following form
\begin{equation}
\label{526}
\hat{H}_{int} = -\hat{\vec{\mu}}_{{\rm J}} \cdot \vec{H}_{ex} \;.
\end{equation}
\noindent Here the operator of the total magnetic moment $\hat{\vec{\mu}}_{{\rm J}}$ of the system is composed additively of the orbital and spin parts
\begin{equation}
\label{527}
\hat{\vec{\mu}}_{{\rm J}} = \hat{\vec{\mu}}_{{\rm L}} + \hat{\vec{\mu}}_{{\rm S}} \;,
\end{equation}
\noindent which themselves are proportional to the corresponding angular momentum operators $\hat{\vec{L}}$ and $\hat{\vec{S}}$ resp., i.e.
\begin{subequations}
\begin{gather}
\label{528a}
\hat{\vec{\mu}}_{{\rm L}} = - \frac{\mu_{{\rm B}}}{\hbar} \hat{\vec{L}} \\
\label{528b}
\hat{\vec{\mu}}_{{\rm S}} = - 2 \frac{\mu_{{\rm B}}}{\hbar} \hat{\vec{S}} \\
\nonumber
\left( \mu_{{\rm B}} \doteqdot \frac{e_{*} \hbar}{2Mc} \;,\text{Bohr magneton} \right) \;.
\end{gather}
\end{subequations}
\noindent Thus, e.g. when the Russell\bs Saunders coupling occurs, the total magnetic moment $\hat{\vec{\mu}}_{{\rm J}}$ \rf{527} can be written as
\begin{equation}
\label{529}
\hat{\vec{\mu}}_{{\rm J}} = - g_{{\rm J}} \cdot \frac{\mu_{{\rm B}}}{\hbar} \hat{\vec{J}} \;,
\end{equation}
\noindent where the gyromagnetic ratio $g_{{\rm J}}$ is given by the Land\'e g\bs factor. In any case, the interaction Hamiltomian $\hat{H}_{int}$ \rf{526} implies the existence of an interaction energy $\Eint$
\begin{eqnarray}
\label{530}
\Eint &=& <\psi|\hat{H}_{int}|\psi> \\
\nonumber
 &=& \frac{\mu_{{\rm B}}}{\hbar} <\psi|\hat{\vec{L}}|\psi> \cdot \vec{H}_{ex} + 2 \frac{\mu_{{\rm B}}}{\hbar} <\psi|\hat{\vec{S}}|\psi> \cdot \vec{H}_{ex} \;,
\end{eqnarray}
\noindent which is experimentally confirmed very well by atomic spectroscopy. This says however that, when the magnetic interactions are correctly incorperated into the RST dynamics, the non\bs relativistic formula \rf{530} for the external magnetic energy $\Eint$ must be also deducible from our RST results. \\
\indent Indeed, such a deduction can easily be attained in the following way: restricting ourselves for a moment to a single particle \kl$a=1$\kr\;with normalized four\bs current $\kmor$ \rf{326}, its external {\it magnetic} energy \kl$ M_{*}^{(m)} c^2$, say\kr\;is deduced either directly from the one\bs particle equation \rf{34a} with $A^2_{\;\;\mu}=B_{\mu}\equiv 0$, or from the one\bs particle version of the energy functional $\ET$ \rf{339}:
\begin{equation}
\label{530n}
\ET \Rightarrow \Ms c^2 + \Emes \;.
\end{equation}
\noindent Indeed, if one could omit here the external term $\Emes$ \rf{248} \kl see below\kr, so that the field energy $\ET$ equals the mass energy $\Ms c^2$~\cite{Hu}, both methods would yield the same result:
\begin{equation}
\label{531}
M_{*}^{(m)} c^2 = \hbar c \vint \: \vec{k}\vvr \cdot \vec{A}_{ex}\vvr \;.
\end{equation}
This is exactly the way in which the considered particle would also magnetically interact
with the other \kl not considered\kr\;particle, cf. \rf{49a}\bs\rf{49b}, so that there is
actually no difference between external and internal magnetism. However the presumed
omission of that external term $\Emes$ in equation \rf{530n} is actually justified for the
presence of a {\it homogeneous} external field $\vec{H}_{ex}$. The reason for this is that
in this case the magnetic volume integral in equ. \rf{248} may be converted to an \kl ill\bs defined\kr\;2\bs surface integral at spatial infinity \kl$r\to\infty\kr$. This surface integral has to be conceived as the \kl infinite\kr\;energy content due to the \kl infinite\kr\;external source emitting the homogeneous field $\vec{H}_{ex}$; and therefore that external term $\Emes$ is to be omitted when one considers the energy of a localized particle. \\ 
\indent Now the vector potential $\vec{A}_{ex}\vvr$ \rf{525} due to a constant magnetic
field $\vec{H}_{ex}$ is given (apart from a gauge transformation) by
\begin{equation}
\label{532}
\vec{A}_{ex}\vvr = - \frac{1}{2} \left( \vec{r} \times \vec{H}_{ex} \right)\;,
\end{equation}
\noindent and therefore the external magnetic energy \rf{531} is found to be of the following form
\begin{equation}
\label{533}
M_{*}^{(m)} c^2 = - \vec{\mu}_{{\rm J}} \cdot \vec{H}_{ex}^{(ph)} \;.
\end{equation}
\noindent Here the magnetic moment $\vec{\mu}_{{\rm J}}$ is given by
\begin{equation}
\label{534}
\vec{\mu}_{{\rm J}} = - \frac{1}{2} e_{*} \vint \left( \vec{r} \times \vec{k}\vvr \right) \;,
\end{equation}
\noindent where $e_{*}$ denotes the elementary charge unit and the physical field $\vec{H}_{ex}^{(ph)}$ is related to our geometric notation $\vec{H}_{ex}$ through
\begin{equation}
\label{535}
\vec{H}_{ex} = \frac{e_{*}}{\hbar c} \vec{H}_{ex}^{(ph)} \;.
\end{equation}
\indent Next, recall that the Gordon decomposition of the four\bs current ${}^{(D}\!j_{\mu}$ of a single Dirac particle reads~\cite{Hu}
\begin{equation}
\label{536}
k_{\mu} \equiv {}^{(D)}\!j_{\mu} = \bar{\psi} \gamma_{\mu} \psi = \frac{i \hbar}{2Mc} \left[ \bar{\psi} \left(\partial_{\mu} \psi \right) - \left( \partial_{\mu}\bar{\psi}\right) \psi \right] + \frac{i \hbar}{Mc} \partial_{\nu} \left( \bar{\psi} \sigma_{\mu}^{\;\;\, \nu} \psi \right) \;,
\end{equation}
\noindent where the objects $\sigma_{\mu \nu}$ are the $\text{Spin}(1,3)$ generators, i.e.
\begin{equation}
\label{537}
\sigma_{\mu \nu} = \frac{1}{4} \left[ \gamma_{\mu} , \gamma_{\nu} \right] \;.
\end{equation}
\noindent Observe here that we are allowed to resort to the Dirac current \rf{536} of a
{\it free} particle, because the magnetic energy \rf{533} already contains the external
field $\vec{H}_{ex}$ linearly and we are satisfied with the deduction of that first\bs
order approximation \rf{530}. Obviously, the Gordon decomposition \rf{536} splits up the
Dirac current $\Djm$ into a sum of two parts, namely the drift part $\dkm$
\begin{equation}
\label{538}
\dkm \doteqdot \frac{i \hbar}{2Mc} \left[ \bp \left( \pmu \psi \right) - \left( \pmu \bp \right) \psi \right]
\end{equation}
\noindent and the polarization part $\pkm$
\begin{equation}
\label{539}
\pkm = \frac{i \hbar}{Mc} \pnu \left( \bp \sigma_{\mu}^{\;\;\, \nu} \psi \right) \;,
\end{equation}
\noindent where both parts obey a separate continuity equation, i.e. for the free particle:
\begin{subequations}
\begin{align}
\label{540a}
\pmo \dkm &= 0 \\
\label{540b}
\pmo \pkm &= 0 \;.
\end{align}
\end{subequations}
For the Gordon decomposition of a composite system see ref. \cite{Hu}). Now insert the splitting
\begin{equation}
\label{541}
\kmu = \dkm + \pkm
\end{equation}
\noindent into the definition of the total magnetic moment $\vmJ$ \rf{534} and find its analogous splitting as
\begin{equation}
\label{542}
\vmJ = \vmL + \vmS \;,
\end{equation}
with the orbital part $\vmL$ being given by
\begin{equation}
\label{543a}
\vmL = - \frac{1}{2} e_{*} \vint \: \left( \vec{r} \times \kdr \right) \;, \\
\end{equation}
\noindent and similarly for the spin part $\vmS$
\begin{gather}
\label{543b}
\vmS = - \frac{1}{2} e_{*} \vint \: \left( \vec{r} \times \kpr \right) \\
\nonumber
\left( \kdr \doteqdot \left\{ {}^{(d)}\!k^j \right\} , \kpr \doteqdot \left\{ {}^{(p)}\!k^j \right\} \right) \;.
\end{gather}
\indent But clearly, the splitting of the total magnetic moment $\vmJ$ \rf{542} induces an analogous splitting of the magnetic energy \rf{533}, i.e.
\begin{equation}
\label{544}
M_{*}^{(m)} c^2 = \EL +  \ES \;,
\end{equation}
\noindent with the orbital part $\EL$ being found as
\begin{equation}
\label{545}
\EL = - \vmL \cdot \Hex = \frac{1}{2} e_{*} \Hex \cdot \vint \: \left( \vec{r} \times \kdr \right) \;,
\end{equation}
\noindent and similarly for the spin part $\ES$
\begin{equation}
\label{546}
\ES = - \vmS \cdot \Hex = \frac{1}{2} e_{*} \Hex \vint \: \left( \vec{r} \times \kpr \right) \;.
\end{equation}
\indent Now it appears as a matter of course that when the RST correction $M_{*}^{(m)}
c^2$ \rf{544} is to be identified with the external magnetic energy $\Eint$ \rf{530},
the orbital part $\EL$ \rf{545} must be identified with the first part of $\Eint$, i.e.
\begin{equation}
\label{547}
\EL \Rightarrow \frac{\mB}{\hbar} \Hex \cdot \vint \: \left( \bpr \cdot \hat{\vec{L}} \cdot \pr \right) \;,
\end{equation}
\noindent and similarly the spin part $\ES$ \rf{546} with the second part of $\Eint$:
\begin{equation}
\label{548}
\ES \Rightarrow 2 \frac{\mB}{\hbar} \Hex \cdot \vint \: \left( \bpr \cdot \hat{\vec{S}} \cdot \pr \right) \;. \\
\end{equation}
\noindent Indeed, this claim \rf{547}\bs \rf{548} can easily be verified: first consider the orbital part $\EL$ \rf{545} and observe that the drift current $\kdr$ reads in three\bs vector notation \kl to be deduced from its four\bs vector version \rf{538}\kr
\begin{equation}
\label{549}
\kdr = \frac{1}{2Mc} \left[ \bp \cdot \left(\frac{\hbar}{i} \nv \psi \right) - \left(\frac{\hbar}{i} \nv \bp \right) \cdot \psi \right] \;.
\end{equation}
\noindent Substituting this into the definition of the orbital magnetic moment $\vmL$ \rf{543a} and observing the Hermiticity of the angular momentum operator $\hat{\vec{L}}$ yields immediately
\begin{equation}
\label{550}
\vmL = - \frac{e_{*}}{2Mc} \vint \: \left( \bp \cdot \hat{\vec{L}} \cdot \psi \right) \;.
\end{equation}
\noindent But with this result, the orbital magnetic energy $\EL$ \rf{545} is easily seen to coincide with its expected form \rf{547} and thus the orbital part of the magnetic interaction energy $\Eint$ \rf{530} is in perfect agreement with RST. \\
\indent By a similar reasoning it is also possible to verify the expected correspondence  of the spin part \rf{548}, albeit by means of a somewhat more subtle argument. For this purpose, observe first that the space part $\vec{k}_{p}$ of the spin polarization current $\pkm$ \rf{539} reduces to a three\bs curl if the stationary field configurations \rf{31a}\bs\rf{31b} are considered:
\begin{equation}
\label{551}
\vec{k}_p\vvr = \nv \times \vec{P}_S\vvr  \;.
\end{equation}
\noindent Here the spin polarization density $\vec{P}_S\vvr$ is defined by
\begin{equation}
\label{552}
\vec{P}_S\vvr = \frac{1}{Mc} \bpr \cdot \hat{\vec{S}} \cdot \pr \;,
\end{equation}
\noindent where the spin\bs operator $\hat{\vec{S}}$ is introduced through
\begin{gather}
\label{553}
\hat{\vec{S}} = \frac{\hbar}{2} \vec{\sigma} \\
\nonumber
\left( \vec{\sigma} = \left\{ \sigma^j \right\} \;; \sigma^{j k} = \frac{i}{2} \varepsilon^{k j}_{\;\;\,\,l}\sigma^l \right) \;.
\end{gather}
\noindent Now, when that polarization current $\vec{k}_p\vvr$ \rf{551} is inserted into the definition of the spin magnetic moment $\vmS$ \rf{543b}, one ends up with the following final form \kl by means of integrating by parts\kr:
\begin{equation}
\label{554}
\vmS = - \frac{e_{*}}{Mc} \vint \: \bp \cdot \hat{\vec{S}} \cdot \psi \;.
\end{equation}
\noindent But clearly, with this result the RST spin magnetic energy $\ES$ \rf{546} is again identified with the spin magnetic part of the interaction energy $\Eint$ \rf{530} and this verifies the claimed correspondence \rf{548}. \\
\indent Summarizing, the RST proposal \rf{531} for the external magnetic interaction energy is found to be in full agreement with the non\bs relativistic description of the \kl experimentally well\bs established\kr\: Zeeman effects; and this in turn may be taken as confirmation that also the internal magnetic interparticle interactions are correctly taken into account by the RST energy functional $\ET$ \rf{339}.


\section{Ground\bs State Interaction Energy}

For a demonstration of the results obtained so far, one may choose the simplest member of the para\bs system, which is the two\bs electron ground\bs state in the Coulomb field \rf{39}. Clearly, the arguments do apply also to all those excited states which own the same symmetry as the two particle ground\bs state, i.e. the highest possible symmetry. The simplification originates here from the fact that for the electrostatic approximation the mass eigenvalues $\tilde{M}_a$ and the spatial parts of the wave functions $\aRpm$ become identical, see equations \rf{315a}\bs\rf{318}; and furthermore the time components $B_0\vvr$ and $\ho$ of the exchange vector potential $B_{\mu}$ and exchange current $h_{\mu}$ do vanish, which is consistent with the Poisson equation \rf{37b}. Furthermore, the identity of both mass eigenvalues implies that the exchange length $\aM$ \rf{422} becomes infinite which then annihilates the ``electric'' exchange field $\vec{X}\vvr$ \rf{423a} \kl$ \leadsto \X \equiv 0$\kr. As a consequence there are no ``electric'' exchange corrections \kl$\Max = 0$, see equation \rf{426}\kr, and thus the ``magnetic'' exchange corrections \rf{427} simplify to
\begin{equation}
\label{51}
\hat{z}_1 \cdot \Meg c^2 = \hat{z}_2 \cdot \Mzg c^2 = \frac{\hbar c}{4 \pi \as} \vint\; \Ys \cdot \Y \;.
\end{equation}
\indent However, since the ``magnetic'' exchange field $\Y$ and its vector potential $\Bv$ \rf{423b} do persist, they give rise to the emergence of exchange corrections for the para\bs system which are beyond the electrostatic approximation. The exchange vector potential $\Bv$ obeys now the ordinary Poisson equation \kl cf. equ. \rf{435}\kr
\begin{equation}
\label{52}
\Delta \Bv = - 4\pi\as \hvs \;.
\end{equation}
\noindent with the corresponding simplification of the solution \rf{438}
\begin{equation}
\label{53}
\Bv = \as \vints \frac{\hvsp}{\brrs}  \;.
\end{equation}
\noindent Observe also that the exchange three\bs current $\hv$ must become sourceless on behalf of the vanishing of its time component $\ho$ \rf{436}
\begin{equation}
\label{54}
\nv \cdot \hvs = 0 \;.
\end{equation}
\noindent Thus the Maxwellian exchange system \rf{439a}\bs\rf{439d} becomes truncated to its ``magnetic'' part
\begin{subequations}
\begin{align}
\label{55a}
\nv \cdot \Y &= 0 \\
\label{55b}
\nv \times \Y &= 4\pi\as \cdot \hvs \;.
\end{align}
\end{subequations}
\indent Summarizing, the lowest\bs order corrections beyond the electrostatic approximation are described by the magnetostatic fields $\vec{H}_a\vvr$ \rf{420} and the ``magnetic'' exchange field $\Y$  \rf{51} so that the total correction $\Delta \ET^{(mg)}$ \rf{457} becomes for the ground\bs state of the para\bs system
\begin{eqnarray}
\label{56}
\Delta \ET^{(mg)} &\Longrightarrow& \frac{1}{2} \sum_{a=1}^{2} \hat{z}_a \cdot \Mamg c^2 \\
\nonumber
 &=& \frac{1}{2} \hat{z}_1 \cdot \left( \Mem + \Meg \right) c^2 + \frac{1}{2} \hat{z}_2 \cdot \left( \Mzm + \Mzg \right) c^2 \;.
\end{eqnarray}
\noindent Since this energy correction consists of the proper magnetostatic contributions \kl$\Mam$\kr and the ``magnetic'' exchange contributions \kl$\Mag$\kr, both subsystems have to be inspected now separately.

\subsection{Exchange Corrections $\Mag$}

Obviously, it is merely a technical question whether one prefers to compute the exchange masses $\Mag$ in their original form \rf{411c}\bs\rf{411d}, being based upon the simultaneous use of the exchange current $\hv$ and the exchange vector potential $\Bv$, or in the form \rf{444a}\bs\rf{444b} which makes use of the currents alone \kl$a=1,2$\kr
\begin{equation}
\label{57}
\hat{z}_a \cdot \Mag c^2 = e^2 \iint d^3\vec{r}\, d^3\vec{r}\,' \: \frac{\hv \cdot \hvsp}{\brrs} \;,
\end{equation}
\noindent or whether one prefers to deal with the above form \rf{51} relying exclusively upon the ``magnetic'' exchange field strenghts $\Y$. In any case one must know explicitly the functional form of the exchange current $\hv$. Therefore one first inserts the general stationary form of the wave functions $\psi_a(\vec{r},t)$ \rf{31a}\bs\rf{31b} into the general definition of the exchange current $h_{\mu}(\vec{r})$ \rf{38b} and then finds the following form for the three\bs current $\hv$:
\begin{equation}
\label{58}
\hv = \frac{i}{4\pi}{\mathbb R}_{+}(r) \cdot \Wps \;.
\end{equation}
\noindent Here the three\bs vector $\vec{W}_p$ depends on the spherical polar coordinates $\vartheta, \varphi$ in the following way
\begin{equation}
\label{59}
\Wp = - \cos\vartheta \left( \vec{e}_x + i \vec{e}_y \right) + e^{i \varphi} \sin\vartheta\: \vec{e}_z \;,
\end{equation}
\noindent where $\vec{e}_x,\vec{e}_y,\vec{e}_z$ are the basis vectors due to a Cartesian parametrization $\kl x,y,z\kr$ of Euclidian three\bs space. Futhermore, the radial function $\mathbb{R}_{+}$ has been defined in terms of the ansatz functions $\aRpm$ for the Pauli spinors ${}^{(a)}\!\phi_{\pm}$ as shown by equation \rf{331}. Since we are satisfied for the moment with the lowest\bs order approximation for the exchange masses $\Mag$, we can resort to the non\bs relativistic approximation for the function $\mathbb{R}_{+}$ \rf{332}. \\
\indent With the exchange current $\hv$ \rf{58} being at hand now, one can in the next step look for the solution $\B$ \rf{53} of the Poisson equation \rf{52}. Clearly, the desired vector potential $\B$ will have the same symmetry as its source $\hvs$, i.e. one tries the product ansatz
\begin{equation}
\label{514}
\Bv = i\cdot r B(r) \cdot \Wp \;,
\end{equation}
\noindent and then one deduces the following differential equation for the radial ansatz function $B(r)$ from the Poisson equation \rf{52}:
\begin{equation}
\label{515}
\frac{{\rm d}^2 B(r)}{{\rm d}r^2} + \frac{4}{r} \frac{{\rm d}B(r)}{{\rm d}r} = \as \cdot \frac{{\mathbb R}_{+}}{r} \;.
\end{equation}
\noindent The solution of this equation is easily worked out as
\begin{equation}
\label{516}
\begin{split}
B(r) = \frac{B_{*}}{3} \left( \frac{z_{ex}}{a_{{\rm B}}} \right)^3 \left\{ 2 \left[ \left( \frac{z_{ex}r}{a_{{\rm B}}} \right)^{-1} + \left( \frac{z_{ex}r}{a_{{\rm B}}} \right)^{-2} \right] \cdot \exp \left( -2 \frac{z_{ex} r}{a_{{\rm B}}} \right) \right. \\ 
\left. + \left( \frac{z_{ex} r}{a_{{\rm B}}} \right)^{-3} \cdot \left[ \exp \left( -2 \frac{z_{ex} r}{a_{{\rm B}}} \right) -1 \right] \right\} \;. \\
\end{split}
\end{equation}
\noindent Obviously, this solution decays as $r^{-3}$ at infinity \kl$r \to \infty$\kr\;but is regular at the origin \kl$r=0$\kr
\begin{equation}
\label{517}
B(r) = \frac{B_{*}}{3} \left( \frac{z_{ex}}{a_{{\rm B}}} \right)^3 \cdot \left\{ - \frac{4}{3} + 2 \frac{z_{ex}r}{a_{{\rm B}}} + \cdots \right\} \;.
\end{equation}
\noindent The constant $B_{*}$ is related to the normalization constant ${\mathbb N}_{*}$
\rf{327} through (non-relativistic limit)
\begin{equation}
\label{518}
B_{*} =\frac{3}{8}\left( \frac{a_{{\rm B}}}{z_{ex}} \right)^4\cdot\zex\as^2\left(2M N_{*}^2\right) = \frac{3}{2} \as^2 \aB\;.
\end{equation}
\indent But once the exchange vector field $\Bv$ is known, one can consider its curl $\Y$ \rf{423b}, which appears in the following form
\begin{equation}
\label{519}
\Y = - \tilde{B}(r) \left[ \vec{e}_x + i \vec{e}_y \right] + \frac{x + iy}{r} \frac{{\rm d}B(r)}{{\rm d} r} \cdot \vec{r} \;,
\end{equation}
\noindent with the radial function $\tilde{B}(r)$ being given by
\begin{equation}
\label{520}
\tilde{B}(r) \doteqdot r \frac{{\rm d}B(r)}{{\rm d} r} + 2B(r) \;.
\end{equation}
\noindent When this result is used in order to compute the exchange corrections $\Mag$ \rf{51}, one is left after the the angular integration with the following radial problem \kl$a =1,2$\kr: 
\begin{equation}
\label{521}
\hat{z}_a \cdot \Mag c^2 = \frac{4}{3} \frac{\hbar c}{\as} \int\limits_0^{\infty} {\rm d}r \:r^2 \left[\tilde{B}(r)^2 + 2B(r)^2 \right] \;.   
\end{equation}
\noindent However, observing here the specific functional form of $\tilde{B}(r)$ \rf{520} and repeatedly integrating by parts yields
\begin{equation}
\label{522}
\hat{z}_a \cdot \Mag c^2 = - \frac{4}{3} \hbar c  \int\limits_0^{\infty} {\rm d}r \:r^3 B(r) \cdot {\mathbb R}_{+}(r) \;,
\end{equation}
\noindent where the differential equation for $B(r)$ \rf{515} has also been used. Now it is just this latter form \rf{522} for the exhange corrections $\Mag$ which can also be recovered by starting from their original form \rf{411c}\bs\rf{411d} and using hereby the previously found results for the vector potential $\Bv$ \rf{514} and the exchange current $\hv$ \rf{58}. \\
\indent Clearly, the general equivalence of both forms \rf{411c}\bs\rf{411d} and \rf{427} for the exchange corrections $\Mag$ has thus been exemplified merely in the lowest order \kl beyond the electrostatic approximation\kr. But this can just be taken as a successfull consistency test of the applied approximation technique, which consists in a combination of {\it linearizing} the magnetic interactions and additionally taking the {\it non\bs relativistic limit}. Indeed, the remaining radial integration of equation \rf{522} can easily be done by use of the known functional forms of $B(r)$ \rf{516} and ${\mathbb R}_{+}(r)$ \rf{332} which yields
\begin{equation}
\label{523}
\hat{z}_1 \cdot \Meg c^2 = \hat{z}_2 \cdot \Mzg c^2 = \frac{1}{6} \left(z_{ex} \as \right)^2 \cdot \frac{z_{ex} e^2}{a_{{\rm B}}} = \frac{4}{15}\left( z_{ex} \as \right)^2\oERe  \;.
\end{equation}
\noindent This result verifies the expectation that the ``magnetic'' interactions are typically smaller than the ``electric'' ones \rf{349n} by a factor $\left(\zex\as\right)^2$ and therefore may be approximated here by their non\bs relativistic limit. But observe on the other hand that the magnetic corrections \rf{523} vary as $\zex^3$ and therefore will become more important for the heavy atoms in comparison to their electric counterparts $\oERe$ \rf{349n} raising only linearly with $\zex$. 

\subsection{Relativistic Normalization}

It must be remarked that the value of the renormalized ``charges'' $\hat{z}_{a}$ \rf{335} do not enter explicitly the result \rf{523} so that the values of the corresponding  renormalization integrals \rf{335} are not explicitly needed. But clearly, the renormalized charges $\hat{z}_a$ do enter the results in an implicit manner, namely through the proper relativistic normalization conditions \cite{Ve2,Hu}. However, it is easy to demonstrate that for the presumed linearization \rf{412a}\bs\rf{412b} the renormalization of the charges is trivial, i.e. one can identify: $\hat{z}_a = 1$. The reason for this is that the deviation of $\hat{z}_a$ from unity is induced by the integral of the entanglement vector $G_{\mu}$ \cite{Ve,Ve2,Hu} 
\begin{equation}
\label{523n}
G_{\mu} = \frac{i}{4\pi\as}\left[B^{\nu}G_{\nu\mu}^{*} - B^{*\nu}G_{\nu\mu}\right]
\end{equation}
\noindent over the time slices $t = \text{const}$ which gives for the stationary field configurations
\begin{eqnarray}
\label{524}
z_1 - \hat{z}_2 = - \left( z_2 - \hat{z}_2 \right) = \int\limits_{t = \text{const}} G_{\mu} {\rm d} S^{\mu} \Longrightarrow \vint & & \!\!\!\!\! G_0(\vec{r}) \\
\nonumber
 &=& - \frac{1}{2\pi\as\aM} \vint \: \Bvs \cdot \Bv \;.
\end{eqnarray}
\noindent However, for the ground state the exchange length $\aM$ becomes infinite \kl$\aM \to \infty$\kr\: and thus the difference \rf{524} between $z_a$ and $\hat{z}_a$ vanishes: $ z_a = \hat{z}_a = 1 \;, (a = 1,2)$. \\
\indent This coincidence of the charges $z_a$ and $\hat{z}_a$ is not only a byproduct of the applied linearization, but for the ground state it has a deeper origin. Indeed, it is closely related to the symmetry of the two\bs particle ground\bs state. Observe here again that, when the normalization integral is done over a time slice \kl$t = const$\kr\;of the Minkowskian space\bs time, it is only the time component $G_0$ of the entanglement vector $G_{\mu}$ \rf{523n} which is relevant:
\begin{equation}
\label{524a}
G_0 \vvr = \frac{i}{4\pi\as} \left[ \Bvs \cdot \X - \Bv \cdot \Xs \right] \;.
\end{equation}
\noindent This component however vanishes, because the ``electric'' exchange vector $\X$
must be put to zero when its source $h_0\vvr$ vanishes, see equation \rf{439a}\bs\rf{439b}. The latter fact, however, is a consequence of the symmetry of the ground\bs state ansatz \rf{31a}\bs\rf{33f}.     

\subsection{Magnetic Corrections $\Mam$}

In order to complete the magnetic part $\Delta \ET^{(mg)}$ \rf{56} of the interaction energy, we have to consider the magnetostatic mass corrections $\Mam$ \rf{49a}\bs\rf{49b}. This will be done by explicit computation of the magnetostatic vector potentials $\vec{A}_a\vvr$ \rf{431b}, however through solving directly the associated Poisson equation \rf{429b}, instead of computing the integral \rf{431b}. The reason for this is that both magnetostatic potentials $\Ava$ \kl$a=1,2$\kr\;are of a similar functional form as the ``magnetic'' exchange potential $\B$ \rf{514}, so that one can take over the functional form of the corresponding solution for the desired potentials $\Ava$. Clearly, this similarity of the vector potentials $\B$ and $\Ava$ has a deeper geometric meaning to be discussed readily. \\
\indent The close relationship between all three potentials $\B$, $\Ava\;(a=1,2)$ is recognized most immediatly by demonstrating that the Poisson equations for the magnetic potentials $\Ava$ \rf{429b} are effictively the same as for the exchange potential $\B$ \rf{52}, i.e. the right\bs hand sides of both Poisson equations effectively do agree. This does not mean that the currents $\kva$ are identical to the exchange current $\hv$ \rf{58}, but when one computes the three\bs currents $\kva$ \rf{38a} by means of the stationary field configurations \rf{31a}\bs\rf{31b}, one finds the following form:
\begin{equation}
\label{555}
\kva = k_a(r) \cdot \vec{V}_p(\vartheta, \varphi) \;.
\end{equation}
\noindent Here the vector field $\vec{V}_p$ is given by
\begin{equation}
\label{556}
\vec{V}_p(\vartheta, \varphi) = \sin\vartheta \cdot \left[ -\sin\varphi \cdot \vec{e}_x + \cos\varphi \cdot \vec{e}_y \right] \doteqdot \sin\vartheta\cdot\vec{e}_{\varphi}
\end{equation}
\noindent and thus is to be conceived as the magnetic analogue of its exchange counterpart $\vec{W}_p(\vartheta, \varphi)$ \rf{59}. Furthermore, the scalar prefactors $k_a(r)$ in equation \rf{555} are found to be of the following form
\begin{subequations}
\begin{align}
\label{557a}
k_1(r) &= \frac{1}{2\pi} \eRp \cdot \eRm \\
\label{557b}
k_2(r) &= -\frac{1}{2\pi} \zRp \cdot \zRm \;.
\end{align}
\end{subequations}
\noindent However, since the radial functions of both particles do coincide for the ground\bs state \kl i.e. $\eRp = \zRp \;, \eRm = \zRm$\kr\;one has
\begin{equation}
\label{558}
k_1(r) = - k_2(r) \equiv \frac{1}{4\pi}{\mathbb R}_{+}(r)\ ,
\end{equation}
\noindent and therefore both single\bs particle currents $\kva$ \rf{555} flow in opposite directions around the $z$\bs axis \kl see fig.1\kr
\begin{equation}
\label{559}
\kve = - \kvz \doteqdot \kv \;.
\end{equation}
\clearpage
\begin{figure}[t]
\begin{center}
\epsfig{file=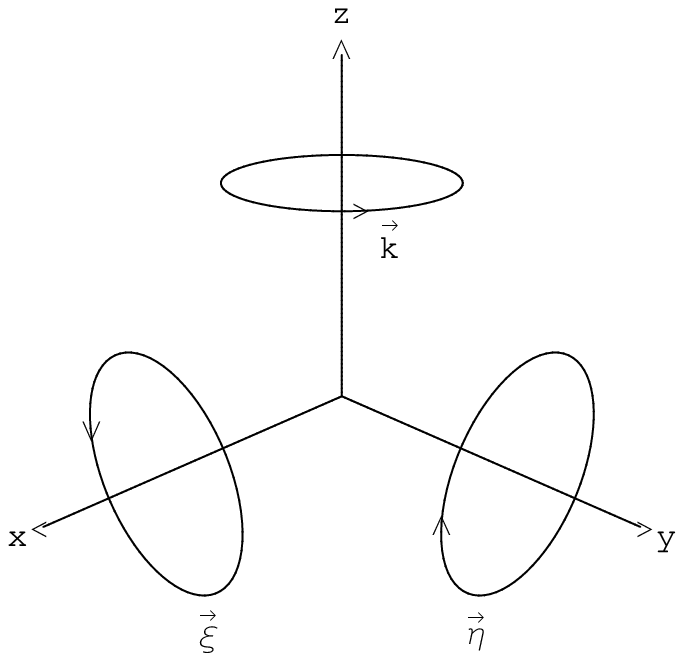}
\end{center}
\vspace{-3cm}
\caption{\label{F1} {\it Isotropy of the Ground\bs State\\} Each coordinate axis is encircled by the same kind of circular flow $\vec{k}$ \rf{559}, $\vec{\xi}$ \rf{560b} and $\vec{\eta}$ \rf{560c} which yields the equipartition of the magnetic interaction energy $\Delta\ETmg$ \rf{570} into the identical contributions $\Delta \ET^{(z)}$ \rf{571}, $\Delta \ET^{(x)}$ \rf{573a} and $\Delta \ET^{(y)}$ \rf{573b}.} 
\end{figure}

Now recall the expectation that the ground\bs state symmetry has to be the highest possible one, so that the $z$\bs axis cannot be singled out in any way as compared to the $x$\bs and $y$\bs axis. Indeed, this is really the case if one takes also into account the exchange current $\hv$ \rf{58} which splits up into its real and imaginary parts $\vec{\xi}\vvr$ and $\vec{\eta}\vvr$ as follows:
\begin{subequations}
\begin{align}
\label{560a}
\hv &= \vec{\xi}\vvr  + i \vec{\eta}\vvr \\
\label{560b}
\vec{\xi}\vvr &= \frac{{\mathbb R}_{+}(r)}{4\pi} \left[ - \cos\vartheta \cdot \vec{e}_y + \sin\vartheta \sin\varphi \cdot \vec{e}_z \right]\\ 
\label{560c}
\vec{\eta}\vvr &= \frac{{\mathbb R}_{+}(r)}{4\pi} \left[ - \cos\vartheta \cdot \vec{e}_x + \sin\vartheta \cos\varphi \cdot \vec{e}_z \right] \;.
\end{align}
\end{subequations}
\noindent Evidently, this result says now that all three axes are encircled by the same
type of flow, namely the $x$\bs axis by $\vec{\xi}\vvr$ \rf{560b}, the $y$\bs axis by
$\vec{\eta}\vvr$ \rf{560c} and the $z$\bs axis by $\kv$ \rf{559}; see fig.1. Here it must
be stressed that this isotropic geometry for the ground\bs state could only be attained by
completing the magnetostatic potentials $\Ava$ by the exchange potential $\B$ into a
triplet of vector potentials. Since $\B$ owes its existence to the use of the non\bs
abelian group $U(2)$ instead of the abelian $U(1)$, which is generally used in the
conventional approach to the electromagnetic interactions  \kl classical and quantum\kr\;,
the ''magnetic'' exchange interactions are now seen to equip the RST with a truly {\it
  non\bs abelian} character!

For the explicit construction of the magnetostatic potentials $\Ava$, one tries an ansatz which has the same symmetry as their sources $\kva$ \rf{555}, quite similarly as it was done for the exchange potential $\B$ \rf{514}, i.e. one puts
\begin{subequations}
\begin{align}
\label{561a}
\Ave &= r A_1(r) \cdot \vec{V}_p(\vartheta, \varphi) \\
\label{561b}
\Avz &= r A_2(r) \cdot \vec{V}_p(\vartheta, \varphi) \;.
\end{align}
\end{subequations}
\noindent Inserting this into the Poisson equations \rf{429b} yields the following differential equations for the radial ansatz functions $A_a(r)$:
\begin{subequations}
\begin{align}
\label{562a}
\frac{{\rm d}^2 A_1(r)}{{\rm d}r^2} + \frac{4}{r} \frac{{\rm d} A_1(r)}{{\rm d}r} &= 2\as \frac{\eRp \cdot \eRm}{r} \equiv 4\pi k_1(r) \\
\label{562b}
\frac{{\rm d}^2 A_2(r)}{{\rm d}r^2} + \frac{4}{r} \frac{{\rm d} A_2(r)}{{\rm d}r} &= 2\as \frac{\zRp \cdot \zRm}{r} \equiv 4\pi k_2(r) \;.
\end{align}
\end{subequations}
\noindent Since for the ground\bs state symmetry the radial parts of both wave functions $\psi_a\vvr$ do agree \kl i.e. $\eRp = \zRp \;, \eRm = \zRm$\kr, one arrives at
\begin{subequations}
\begin{align}
\label{563a}
\frac{{\rm d}^2 A_1(r)}{{\rm d}r^2} + \frac{4}{r} \frac{{\rm d} A_1(r)}{{\rm d}r} &= \as \frac{{\mathbb R}_{+}(r)}{r} \\
\label{563b}
\frac{{\rm d}^2 A_2(r)}{{\rm d}r^2} + \frac{4}{r} \frac{{\rm d} A_2(r)}{{\rm d}r} &= -\as \frac{{\mathbb R}_{+}(r)}{r} \;,
\end{align}
\end{subequations}
\noindent which is effectively the same equation as for the radial exchange function $B(r)$ \rf{515}. Therefore, one can directly take over the solutions as
\begin{equation}
\label{564}
A_1(r) = -A_2(r) \equiv B(r) \;,
\end{equation}
\noindent where the radial function $B(r)$ has already been specified by equation \rf{516}. \\
\indent Clearly, the identification \rf{564} of all radial functions $A_a\vvr$ and $B\vvr$ is a further indication of the ground\bs state isotropy and it is instructive to elucidate this effect also from another point of view. For this purpose, reconsider the source equations \rf{240a}\bs\rf{240d} and insert therein the general shapes of the currents $\kva$ \rf{555} and $\hv$ \rf{58}, together with the vector potentials $\Ava$ \rf{561a}\bs\rf{561b} and $\B$ \rf{514}. The right\bs hand sides of the first two equations \rf{240a}\bs\rf{240b} then turn out to be zero
\begin{equation}
\label{565}
B^{\mu} h_{\mu} - B^{\mu *} h^{*}_{\mu} = 0 \;,
\end{equation}
\noindent and therefore the currents $\kva$ are found to be sourceless
\begin{equation}
\label{566}
\nv \cdot \kva = 0 \;.
\end{equation}
\noindent Observe that this property is also shared by the exchange current $\hv$, cf. \rf{54}. The remaining two equations \rf{240c}\bs\rf{240d} yield the relation
\begin{equation}
\label{567}
B(r) = \frac{A_1(r) - A_2(r)}{k_1(r) - k_2(r)} \cdot \frac{{\mathbb R}_{+}(r)}{4\pi} \;.
\end{equation}
\noindent Now recall here that, because of the ground\bs state symmetry, the denominator becomes on account of equation \rf{558}
\begin{equation}
\label{568}
k_1(r) - k_2(r) = \frac{{\mathbb R}_{+}(r)}{2\pi}
\end{equation}
\noindent and furthermore the numerator appears by virtue of equation \rf{564} as
\begin{equation}
\label{569}
A_1(r) - A_2(r) = 2B(r)
\end{equation}
\noindent so that the relation \rf{567} degenerates to an identity. Therefore the source equations \rf{240a}\bs\rf{240d} are automatically satisfied when all RST objects $\left\{ \kva,\hv,\Ava,\B\right\}$ own the symmetry properties of the ground\bs state, irrespective of the precise functional form of the radial parts $\aRpm$ of the wave functions!\\
\indent With these prerequisites at hand, we are now able to test a further consequence of the ground\bs state isotropy: if it is true that, on account of the isotropic geometry, all three spatial directions are equivalent, then the corresponding currents $\kv$ \rf{559}, $\vec{\xi}\vvr$ \rf{560b} and $\vec{\eta}\vvr$ \rf{560c} have to contribute equal parts to the magnetic energy $\Delta \ET^{(mg)}$ \rf{56}, see fig. 1:
\begin{equation}
\label{570}
\Delta \ET^{(mg)} = \Delta \ET^{(x)} + \ET^{(y)} + \ET^{(z)}\ .
\end{equation}
\noindent Now the contribution of the $z$\bs direction is given by the currents $\kva$, cf. \rf{441}
\begin{equation}
\label{571}
\Delta \ET^{(z)} \doteqdot \frac{1}{2}\left(\hat{z}_1\cdot\Mem c^2 + \hat{z}_2\cdot\Mzm c^2\right) = e^2 \iint d^3\vec{r}\, d^3\vec{r}\,' \frac{\kv \cdot \kvs}{\brrs} \;.
\end{equation}
\noindent Next, the contributions of the $x$\bs and $y$\bs directions are given by the ``magnetic'' exchange energy $\hat{z}_a \cdot \Mag c^2$, see equations \rf{444a}\bs\rf{444b} for infinite exchange length \kl$\aM \to \infty$\kr:
\begin{equation}
\label{572}
\Delta \ET^{(x)} + \Delta \ET^{(y)} \doteqdot \frac{1}{2}\left(\hat{z}_1\cdot\Meg c^2 + \hat{z}_2\cdot\Mzg c^2\right) = e^2 \iint d^3\vec{r}\, d^3\vec{r}\,'\, \frac{\hv \cdot \hvsp}{\brrs} \;,
\end{equation}
\noindent i.e. for either direction separately
\begin{subequations}
\begin{align}
\label{573a}
\Delta \ET^{(x)} &= e^2 \iint d^3\vec{r}\, d^3\vec{r}\,'\, \frac{\vec{\xi}\vvr \cdot \vec{\xi}\vvrs}{\brrs} \\
\label{573b}
\Delta \ET^{(y)} &= e^2 \iint d^3\vec{r}\, d^3\vec{r}\,'\, \frac{\vec{\eta}\vvr \cdot
  \vec{\eta}\vvrs}{\brrs}\ .
\end{align}
\end{subequations}
 But since the three currents $\kv,\vec{\xi}\vvr,\vec{\eta}\vvr$ differ only by their spatial orientation, not by their intrinsic pattern of flux lines and strength, all three contributions are identical, cf. \rf{523}
\begin{equation}
\label{574}
\Delta\ETmg = \Delta \ET^{(x)} + \Delta \ET^{(y)} + \Delta \ET^{(z)} = \frac{1}{4} \left( z_{ex} \as \right)^2 \cdot \frac{z_{ex} e^2}{\aB} =\frac{2}{5} \left( z_{ex} \as \right)^2 \cdot \oERe \;.
\end{equation}
\indent This is the definitive result for the magnetic corrections; and can now to be used in order to check our preliminary estimate of the magnitude of these magnetic contributions  which has been made  within the framework of the electrostatic approximation \kl see end of sect. III\kr. Namely for small values of the coupling constant \kl$\zex\as\ll1$\kr, the preliminary picture of the magnetic interactions did identify the electrostatic RST prediction $\DERSTe$ with the non\bs relativistic interaction energy $\oERe$, see equation \rf{348n}. On the other hand the discrepancy between these RST predictions $\DERSTe$ and the experimental data $\Delta\Eexp$ was attributed to the magnetic energy $\Delta\ETmg$, see equation \rf{356n}. Therefore the limit value $f_0^2$ of the geometric factor $\fs^2$ \rf{357n} for $\left( \zex\as\right)^2 \ll 1$ appears now as
\begin{equation}
\label{643}
f_0^2 = \frac{1}{\left(\zex\as\right)^2}\cdot\frac{\Delta\ETmg}{\oERe} = \frac{2}{5}
\end{equation}
\noindent which appears to be in acceptable agreement with the values of $\fs^2$ as presented in table~I. Clearly, for a more precise test of the limit values $\varepsilon_0$ \rf{351n} and $f_0^2$ \rf{643} one would have to extend table~I to smaller values of the coupling parameter $\zex\as$, possibly up to neutral helium \kl$\zex=2$, see a separate paper\kr.

\subsection{Theory vs. Experiment}

Apart from such an internal consistency test for the RST predictions, it is of course highly interesting to compare the complete RST predictions $\DERSTemg$, i.e the sum of electric and magnetic contributions
\begin{equation}
\label{644}
\DERSTemg \doteqdot \DERSTe + \Delta\ETmg \;,
\end{equation}
\noindent to both the experimental data and to other theoretical predictions, such as the
relativistic 1/Z expansion \cite{Dr}, the multiconfiguration Dirac\bs Fock method \kl MCDF\kr\;\cite{F1}-~\cite{Wi} and relativistic many\bs body pertubation theory \kl MBPT\kr\;\cite{Wi}-\cite{Jo}, or the all\bs order technique for relativistic MBPT \cite{Pl}. The predictions of these four theoretical approaches are collected in table~II together with the experimental data \cite{MES} in order to oppose them to the present RST predictions $\DERSTemg$ \rf{644}.
\begin{table}
\begin{tabular}{c||c|c|c||c|c|c|c}
Element & $\Delta\Eexp$ & $\DERSTemg$ & $\frac{\Delta\Eexp - \DERSTemg}{\Delta\Eexp}$ & Relativistic MBPT & Rel. all order & MCDF & Unified \\
\kl$\zex$\kr &\cite{MES} &\rf{644} & & \cite{Wi,Jo} & MBPT \cite{Pl} & \cite{F1}-\cite{Wi} & \cite{Dr} \\ \hline\hline
Ge \kl32\kr & 562,5$\pm$1,6 & 564,9 & -0,42\% & 561,9 & 562,1 & 562,1 & 562,1 \\ \hline
Xe \kl54\kr & 1027,2$\pm$3,5 & 1031 & -0,37\% & 1028,1 & 1028,4 & 1028,2 & 1028,8 \\ \hline
Dy \kl66\kr & 1341,6$\pm$4,3 & 1336 & 0,41\% & 1336,6 & 1337,2 & 1336,5 & 1338,2 \\ \hline
W \kl74\kr & 1568$\pm$15 & 1570 & -0,11\% & 1574,6 & 1574,8 & 1574,6 & 1576,6 \\ \hline
Bi \kl83\kr & 1876$\pm$14 & 1868 & 0,43\% & 1882,7 & -- & 1880,8 & 1886,3 \\ 
\end{tabular} \medskip\medskip
\caption{\label{T2} Comparison of the theoretical predictions for the ground\bs state interaction energy with the experimental data \cite{MES}. All energies are measured in [eV]. Already after the inclusion of the magnetic interactions in the lowest\bs order approximation, the RST predictions~$\DERSTemg$ \kl third column\kr\;meet with the corresponding predictions of the other theoretical approaches \kl right half of the table\kr.}
\end{table}
As it can easily be seen from a comparison of table~I and table~II, the RST predictions appear now to be very satisfying: after the magnetic interactions have been included into the RST results to yield the interaction energy $\DERSTemg$ \rf{644}, third column of table~II, the deviation from the experimental data $\Delta\Eexp$ \kl second column of table 2\kr\;has decreased to less than a half percent which does no longer depend upon the nuclear charge $\zex$. This means that the discrepancy between the experimental values $\Delta\Eexp$ and the electrostatic RST predictions $\DERSTe$, ranging from 1,7\% for germanium\kl32\kr\;up to 11,5\% for bismuth \kl83\kr\; \kl see table~I\kr, is actually caused by the magnetic interactions with a remaining uncertainty of roughly 0,4\%. This supports the RST picture of the simultaneous action of electric and magnetic forces in the electronic orbits of an atom. \\
\indent A further, very satisfying feature of the RST results $\DERSTemg$ \rf{644} refers
to the comparison to the other four theoretical approaches, see the last four columns of
table~II. Here the RST predictions appear to be of the same order of precision as the
corresponding predictions of the other four theoretical approaches, see table III of
ref.~\cite{MES}. Summarizing, one can therefore judge that the RST predictions can compete
succesfully with these other approaches, even when the magnetic interactions are treated
in the lowest-order approximation!  Clearly, in the next step one will study the higher\bs order approximations of the magnetic RST interactions, which may be expected to lead even closer to the experimental data.



\end{document}